\documentclass[mathematics,article,accept,pdftex,moreauthors]{Definitions/mdpi} 

\firstpage{1} 
\pubvolume{14}
\issuenum{2}
\articlenumber{381}
\pubyear{2026}
\copyrightyear{2026}
\externaleditor{Hanzhou Wu}
\datereceived{7 December 2025} 
\daterevised{18 January 2026} 
\dateaccepted{19 January 2026} 
\datepublished{22 January 2026 } 

\usepackage{subcaption} 

\Title{{Interpreting} 
 Multi-Branch Anti-Spoofing Architectures: Correlating Internal Strategy with Empirical Performance}

\Author{{Ivan Viakhirev} 
 $^{1,2}$\orcidA{}, Kirill Borodin $^{2,}$*\orcidB{}, Mikhail Gorodnichev $^{2}$\orcidC{} and Grach Mkrtchian$^{2,}$*\orcidD{}}

\address{%
$^{1}$ \quad {Faculty of AI technologies, ITMO University}
, {197101 Saint Petersburg, Russia}
; {i.viakhirev@mail.ru}
\\
$^{2}$ \quad {Faculty of IT, Moscow Technical University of Communication and Informatics}, {111024 Moscow, Russia}; {m.g.gorodnichev@mtuci.ru}
}

\corres{Correspondence: k.n.borodin@mtuci.ru (K.B.); g.m.mkrtchyan@mtuci.ru (G.M.)}

\abstract{Multi-branch deep neural networks like AASIST3 achieve state-of-the-art comparable performance in audio anti-spoofing, yet their internal decision dynamics remain opaque compared to traditional input-level saliency methods. While existing interpretability efforts largely focus on visualizing input artifacts, the way individual architectural branches cooperate or compete under different spoofing attacks is not well characterized. This paper develops a framework for interpreting AASIST3 at the component level. Intermediate activations from fourteen branches and global attention modules are modeled with covariance operators whose leading eigenvalues form low-dimensional spectral signatures. These signatures train a CatBoost meta-classifier to generate TreeSHAP-based branch attributions, which we convert into normalized contribution shares and confidence scores ($C_b$) to quantify the model's operational strategy. By analyzing 13 spoofing attacks from the ASVspoof 2019 benchmark, we identify four operational archetypes—ranging from ``Effective Specialization'' (e.g., A09, Equal Error Rate (EER) 0.04\%, $C=1.56$) to ``Ineffective Consensus'' (e.g., A08, EER 3.14\%, $C=0.33$). Crucially, our analysis exposes a ``Flawed Specialization'' mode where the model places high confidence in an incorrect branch, leading to severe performance degradation for attacks A17 and A18 (EER 14.26\% and 28.63\%, respectively). These quantitative findings link internal architectural strategy directly to empirical reliability, highlighting specific structural dependencies that standard performance \mbox{metrics overlook}.
}



\begin{document}

\section{Introduction}
\label{sec:introduction}

The rapid advancement of deep learning has led to significant improvements in synthetic speech generation, including Text-to-Speech (TTS) and Voice Conversion (VC) technologies~\citep{asvspoof2021, asvspoof5}. While beneficial for legitimate applications, these technologies pose a severe security threat to Automatic Speaker Verification (ASV) systems, necessitating robust anti-spoofing countermeasures (CMs)~\cite{wang2020asvspoof2019largescalepublic, Borodin2024ResCapsRes2TCNGuard}. State-of-the-art CMs, such as the AASIST3 architecture~\cite{Borodin2024aasist3}, have evolved into complex, multi-branch neural networks capable of detecting sophisticated spoofing attacks. However, as~the complexity of these models increases, their internal decision-making processes become increasingly opaque. Understanding {how} 
 these models arrive at a decision---specifically, how they utilize their parallel processing branches---is no longer merely a matter of academic interest but an important requirement for ensuring the security and reliability of biometric systems~\citep{tdcf}.

While multi-branch architectures like AASIST3 achieve low Equal Error Rates (EER) on benchmark datasets like ASVspoof 2019~\cite{wang2020asvspoof2019largescalepublic}, the~functional role of each branch remains largely unexplored. It is unclear whether the four parallel branches (B0--B3) and the global attention modules (GAT) function as a redundant ensemble to improve stability or if they develop specialized roles to detect specific spoofing artifacts. The~primary challenge in answering this question lies in the non-linear and high-dimensional nature of deep feature representations. Traditional interpretability methods often focus on input-level saliency maps, which may fail to capture the intermediate coordination strategies employed by the network's internal components~\citep{Integrated_gradients}, such as the Heterogeneous Stacking Graph Attention Layers (HSGAL)~\citep{aasist}.

The field of neural network interpretability has seen the emergence of spectral analysis techniques to probe the internal state of models~\citep{HuSompolinsky2022CovSpectrum}. Recent studies have demonstrated that the eigenvalue spectrum of activation covariance matrices can serve as a robust signature of a model's operational state~\cite{sihag2023covarianceneuralnetworks}. For~instance, Binkowski~et~al.~\cite{binkowski2025hallucination} successfully utilized spectral features of attention maps to detect hallucinations in Large Language Models. Similarly, El Harzli~et~al.~\cite{harzli2025adversarial} showed that adversarial perturbations systematically shift activations toward specific eigenspaces. Parallel to this, game-theoretic approaches like SHAP (SHapley Additive exPlanations) \cite{Lundberg2017unified} have become the standard for fair feature attribution. However, few studies have combined these spectral and game-theoretic methods to analyze the architectural efficiency and vulnerability of multi-branch audio anti-spoofing networks~\citep{ge2024explainingdeeplearningmodels, YU2024103720}.

Despite the availability of these tools, there is a notable gap in the literature regarding the correlation between a model's internal processing strategy and its empirical performance on specific attack types~\citep{li2025surveyspeechdeepfakedetection}. Current research often treats the model as a monolithic entity, ignoring the potential for internal conflicts or misplaced confidence within specific architectural branches~\citep{Pomponi2021ProbabilisticConfidenceMultiExit}. There is a lack of systematic methodologies that can quantify whether a model is succeeding due to a robust consensus among its components or failing due to an over-reliance on a single, incorrect feature extractor~\citep{heidemann2021measuring}. Addressing this gap is essential for diagnosing high-error failure modes and designing more resilient~architectures.

In this work, we propose a novel framework for deconstructing the decision-making process of the AASIST3 model. Our contributions are as~follows:
\begin{itemize}
    \item We introduce a robust feature extraction pipeline based on the spectral analysis of layer activation covariance matrices, covering 14 key internal components of the \mbox{AASIST3 architecture}.
    \item We develop a meta-classification and attribution methodology using {CatBoost v1.2.8} 
 and {TreeSHAP v0.3.1}~\cite{lundberg2019explainableaitreeslocal} to quantify the contribution share of each processing branch (B0--B3) and global module (GAT-S, GAT-T).
    \item We define four operational archetypes---Effective Specialization, Effective Consensus, Ineffective Consensus, and~Flawed Specialization---to classify the model's behavior.
    \item We empirically demonstrate that AASIST3 dynamically adapts its strategy for different attacks (A07--A19) and identify a structural misalignment where the model confidently relies on an incorrect branch, leading to  high error rates.
\end{itemize}

{To address these challenges, this paper is organized as follows. Section~\ref{sec:materials} describes the AASIST3 architecture, the~proposed  interpretation pipeline, and~the confidence-based formulation of branch contribution shares. Section~\ref{sec:results} presents the experimental results, including ablation studies, the~definition and empirical characterization of the four operational archetypes, and~detailed per-attack analyses of internal model behavior. Section~\ref{sec:discussion} discusses the architectural implications of these findings, highlighting robustness, vulnerabilities, and~the role of global attention modules, while Section~\ref{sec:conclusion} concludes the study and outlines directions for future work.
}

\section{Materials and~Methods}\label{sec:materials}
\unskip

\subsection{AASIST3 Architecture~Components}
The AASIST3 architecture~\cite{Borodin2024aasist3} is a sophisticated multi-branch neural network designed for audio anti-spoofing. Our analysis explicitly focuses on interpreting the roles of its 14 primary internal components. These components process the latent representation generated by the initial RawNet2-based~encoder.

The analyzed components are categorized into three functional {groups:} 

\begin{enumerate}
 \item[(1)] Heterogeneous Stacking Graph Attention Layers (HSGAL): These layers form the computational core of the parallel branches. They employ graph attention mechanisms to capture complex, non-local spectro-temporal patterns within the audio data. We analyze the early-stage (HSGAL1) and late-stage (HSGAL2) layers across all four branches: B0-HSGAL1, B0-HSGAL2, B1-HSGAL1, B1-HSGAL2, B2-HSGAL1, B2-HSGAL2, and~B3-HSGAL1, B3-HSGAL2.
 \item[(2)] Pooling Layers (Pool): Each branch includes a pooling operation for feature aggregation and dimensionality reduction. We analyze these as: B0-Pool, B1-Pool, B2-Pool, and~B3-Pool.
  \item[(3)] Global Graph Attention Networks (GAT): Two global modules operate on the multidimensional features to capture holistic dependencies. GAT-S (Spectral) models relationships across frequency bins, while GAT-T (Temporal) models dependencies across time~frames.
\end{enumerate}

\subsection{Methodology~Pipeline}
The proposed analysis framework follows a three-phase pipeline designed to extract robust spectral features and quantify the contribution of each AASIST3~\cite{Borodin2024aasist3} component.

\begin{figure}[H]
    \includegraphics[width=.99\linewidth]{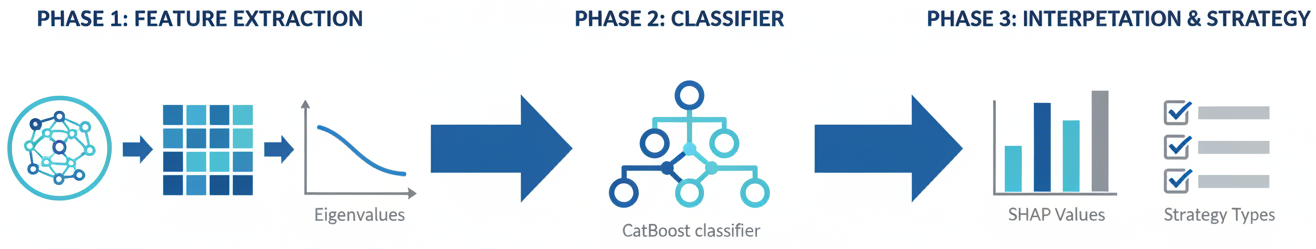}
    \caption{{Schematic} 
 overview of the proposed Spectral–SHAP interpretation~pipeline.}
    \label{fig:pipeline}
\end{figure}

{Figure~\ref{fig:pipeline} provides a high-level visual summary of our three-phase analysis framework. In~Phase 1, raw intermediate activations from the AASIST3 encoder are compressed into robust spectral signatures ($\lambda_1, \cdots, \lambda_10$) that capture the principal variations of the feature space. These signatures serve as the input for Phase 2, where a gradient-boosted decision tree (CatBoost) learns to map spectral patterns to specific spoofing attacks. Finally, in~Phase 3, we leverage the trained meta-classifier to extract Shapley values, which are aggregated into confidence scores and operational strategy classifications, enabling a direct link between internal model behavior and empirical performance.
}

 {Phase 1: Spectral Feature Extraction.} For every audio sample in the evaluation subset of the ASVspoof 2019 Logical Access (LA) dataset~\cite{wang2020asvspoof2019largescalepublic}, we extract intermediate activations from the 14 specified components. Let \( \mathbf{A}^{(l)} \in \mathbb{R}^{D \times N} \) denote the activation matrix for layer \( l \), where \( D \) represents the feature dimension (number of neurons or channels) and \( N \) represents the spatial/temporal dimension (or batch size in batch-wise processing). First, we center the data by subtracting the mean vector \( \boldsymbol{\mu}^{(l)} \), {Equation~(\ref{eq:centring}).} 

\begin{equation}\label{eq:centring}
    \bar{\mathbf{A}}^{(l)} = \mathbf{A}^{(l)} - \boldsymbol{\mu}^{(l)}
\end{equation}
{Next}
, we compute the empirical covariance matrix \( \mathbf{C}^{(l)} \in \mathbb{R}^{D \times D} \), Equation~(\ref{eq:cov_matrix}).
\begin{equation}\label{eq:cov_matrix}
    \mathbf{C}^{(l)} = \frac{1}{N-1} \bar{\mathbf{A}}^{(l)} (\bar{\mathbf{A}}^{(l)})^T
\end{equation}
{To} capture the principal modes of variation within the layer's representation, we perform an eigenvalue decomposition, Equation~(\ref{eq:eig_dec}).
\begin{equation}\label{eq:eig_dec}
    \mathbf{C}^{(l)} \mathbf{v}_k = \lambda_k \mathbf{v}_k
\end{equation}
where \( \lambda_k \) are the eigenvalues and \( \mathbf{v}_k \) are the eigenvectors. The~top-10 eigenvalues \( (\lambda_1^{(l)}, \dots, \lambda_{10}^{(l)}) \) are selected to form the spectral signature for layer \( l \). {In Section~\ref{sec:eigenvalues_ablations}, we conduct ablation studies by varying the number of selected eigenvalues used to construct the spectral signature for each layer.}

 {Phase 2: Meta-Classification.} The spectral signatures from all 14 layers are concatenated to form a meta-feature vector. This vector is used to train a CatBoost classifier~\cite{Hajjouz2025enhancing}. The~model is trained on a multi-class classification task where targets are specific attack types (e.g., A07, A08). Tree-based models like CatBoost are chosen for their compatibility with TreeSHAP~\cite{lundberg2019explainableaitreeslocal}, allowing for exact and efficient computation of Shapley~values.

{The CatBoost Classifier was initialized with the default MultiClass objective, which optimizes the Multinomial Cross-Entropy loss using the softmax function to model probability distributions across the 13 attack classes. We set the number of boosting iterations to 1000 to ensure convergence, while maintaining the default learning rate of 0.03 for the MultiClass. To~control model complexity and prevent overfitting on the spectral meta-features, we utilized the default symmetric tree structure with a maximum depth of 6, and~optimization was performed using 8 parallel threads on a CPU backend.}

 {Phase 3: Strategy Quantification.} Using TreeSHAP~\cite{lundberg2019explainableaitreeslocal}, we compute the SHAP values \( \phi_{i} \) for each of the 14 analyzed architectural components \( i \) and each prediction. We employ two aggregation methods to interpret these values:

{\textit{{Branch Attribution Sum.}} We aggregate the raw SHAP values to determine the total directional contribution of each branch $b$. Let $\Phi_b$ denote the Branch Attribution Sum, defined as the summation of the mean SHAP values of its constituent components, Equation~(\ref{eq:phi})
\begin{equation}\label{eq:phi}
    \Phi_{b} = \sum_{i \in b} \text{mean}(\phi_i)
\end{equation}
where positive $\Phi_b$ indicates that the branch generally pushes the prediction toward the ``spoof'' class, while a negative value indicates a tendency toward ``bona fide.''}

\textit{{Normalized Contribution Share.}} To compare the relative reliance on different branches, we define a confidence score that penalizes internal disagreement within a branch. If~components within a branch (e.g., HSGAL1 vs. Pool) have high variance in their SHAP values, the~branch's overall confidence is reduced. Considering the standard deviation \( \text{std}(\cdot) \) of the mean SHAP values across components within branch \( b \), the~confidence score \( \mathcal{C}_b \) is defined as in Equation~(\ref{eq:confidence_score}).
\begin{equation}\label{eq:confidence_score}
    \mathcal{C}_b = \frac{|\sum_{i \in b} \text{mean}(\phi_i)|}{1 + \text{std}_{i \in b}(\text{mean}(\phi_i))}
\end{equation}
Here, the~denominator \( 1 + \text{std}(\dots) \) acts as a regularization term: branches with consistent component contributions are favored over those with high internal conflict. Finally, we apply the Softmax function to these scores to obtain the percentage share \( S_c \) for each of the six main architectural blocks (B0--B3, GAT-S, GAT-T), Equation~(\ref{eq:dominant_shate}).
\begin{equation}\label{eq:dominant_shate}
    S_c = \frac{e^{\mathcal{C}_c}}{\sum_{j=1}^{6} e^{\mathcal{C}_j}}
\end{equation}

{To validate the robustness of the confidence score formulation in Equation~(\ref{eq:confidence_score}), we perform an ablation study comparing the proposed linear penalty against two alternative formulations: a quadratic penalty $C_b^{quad}$ and an exponential penalty $C_b^{exp}$, as~well as a baseline with no penalty $C_b^{none}$. These are defined as follows, Equation~(\ref{eq:penalties}):}\vspace{-6pt}
\begin{adjustwidth}{-\extralength}{0cm}
\centering 
\begin{equation}\label{eq:penalties}
    C_b^{quad} = \frac{|\sum_{i\in b} \text{mean}(\phi_i)|}{1+\text{std}_{i \in b}(\text{mean}(\phi_i))^2}, \quad C_b^{exp} = \frac{|\sum_{i\in b} \text{mean}(\phi_i)|}{\exp(\text{std}_{i \in b}(\text{mean}(\phi_i)))}, \quad
     C_b^{none} = {|\sum_{i\in b} \text{mean}(\phi_i)|}
\end{equation}
\end{adjustwidth}

{We evaluate the stability of the branch ranking across these methods using the Kendall rank correlation coefficient $\tau$. This analysis aims to demonstrate that while the choice of penalty function modulates the absolute confidence scores, the~identification of the dominant strategic branch remains consistent across different penalty intensities}

 {Operational Archetype Definitions:}
Based on the calculated shares and their correlation with the empirical performance results presented in Section~\ref{sec:results}, we formally classify the model's behavior into four distinct~archetypes.

 {Effective Specialization:} Defined by a low Equal Error Rate (typically \( < 1\% \)) and a high dominant contribution share. The~model successfully identifies a unique artifact and delegates detection to a specialized~component.

 {Effective Consensus:} Defined by a low EER (typically \( < 1\% \)) and a balanced distribution of shares. The~model achieves high performance through a fault-tolerant, distributed decision-making~process.

 {Ineffective Consensus:} Defined by a moderate to high EER (typically \( > 1\% \)) and balanced but weak contribution shares. This indicates model confusion, where no component finds a strong~signal.

 {Flawed Specialization:} Defined by a high EER (typically \( > 10\% \)) despite a high dominant share. This represents a high-error mode where the model is ``confidently wrong,'' relying on a single branch that fails to correctly classify the~attack.

\subsubsection*{ing{Confidence} 
 Interval Estimation for Reported~Metrics}
{For the scalar quantities reported in {Tables}
~\ref{tab:main_analysis} (and summarized in {Table}~\ref{tab:dominant_share_by_archetype}),~\ref{tab:appendix_shap_sum} and~\ref{tab:appendix_softmax} (Equal Error Rates, summed mean SHAP values, and~Softmax-normalized contribution shares), we provide two-sided $(1-\alpha)\times 100\%$ confidence intervals around the sample mean. Let $x_1,\dots,x_n$ denote the values of a given metric computed over $n$ evaluation samples (e.g., utterances) for a fixed attack and component. The~sample mean and sample standard deviation are defined as in Equation~(\ref{eq:mean_std})
\begin{equation}\label{eq:mean_std}
    \bar{x} = \frac{1}{n}\sum_{i=1}^{n} x_i,\qquad
s = \sqrt{\frac{1}{n-1}\sum_{i=1}^{n} (x_i - \bar{x})^2}.
\end{equation}
The standard error of the mean is defined in Equation~(\ref{eq:se}).
\begin{equation}\label{eq:se}
    \mathrm{SE} = \frac{s}{\sqrt{n}}.
\end{equation}
Assuming approximate normality of the sampling distribution of $\bar{x}$, the~two-sided \mbox{$(1-\alpha)\times 100\%$} confidence interval is given by Equation~(\ref{eq:ci}).

\begin{equation}\label{eq:ci}
    \bar{x} \pm z_{1-\alpha/2}\,\mathrm{SE},
\end{equation}
where $z_{1-\alpha/2}$ is the $(1-\alpha/2)$ quantile of the standard normal distribution (for a $95\%$ confidence level, $z_{0.975} \approx 1.96$). In~all tables, we report each estimate in the compact form $\bar{x} \pm \mathrm{CI}$, where $\mathrm{CI} = z_{1-\alpha/2}\,\mathrm{SE}$ denotes the half-width of the corresponding \mbox{confidence interval}.}

\begin{table}[H]
\centering
\caption{{Synthesis} 
 of Model Performance (EER) and Internal Strategy (Contribution Share). Dominant Share refers to the highest single-component share after Softmax normalization. The~reported values include 95\% confidence intervals, expressed in the form value $\pm$ CI, to~reflect the statistical uncertainty of both EER and SHAP-based contribution estimates.
}
\label{tab:main_analysis}
\small
\begin{adjustwidth}{-\extralength}{0cm}
\centering 
\begin{tabularx}{\fulllength}{lCLCCl}
\toprule
\textbf{Attack} & \textbf{EER (\%)} & \textbf{Dominant Component(s)} & \textbf{Dominant \mbox{Share (\%)}} & \textbf{Confidence Score} & \textbf{Identified Strategy} \\ \midrule
A09 & $0.05 \pm 0.01$ & {B2, B1} & $22.85 \pm 0.13$, $21.85 \pm 0.13$ & $2.30 \pm 0.01$, $2.26 \pm 0.01$ & Effective Specialization \\
A14 & $0.27 \pm 0.05$ & {B2, B0} & $26.22 \pm 0.10$, $18.48 \pm 0.08$ & $1.87 \pm 0.00$, $1.52 \pm 0.00$ & Effective Specialization \\
A07 & $0.40 \pm 0.06$ & {B2, B1} & $22.62 \pm 0.11$, $18.84 \pm 0.07$ & $1.68 \pm 0.00$, $1.50 \pm 0.00$ & Effective Specialization \\
A11 & $0.67 \pm 0.07$ & {B0, B2} & $19.50 \pm 0.07$, $19.16 \pm 0.07$ & $1.40 \pm 0.00$, $1.38 \pm 0.00$ & Effective Consensus \\
A16 & $0.74 \pm 0.07$ & {B2, B1} & $19.82 \pm 0.05$, $19.19 \pm 0.05$ & $1.32 \pm 0.00$, $1.29 \pm 0.00$ & Effective Consensus \\
A19 & $0.97 \pm 0.10$ & {B1, B2} & $20.09 \pm 0.07$, $20.02 \pm 0.09$ & $1.45 \pm 0.01$, $1.45 \pm 0.01$ & Effective Specialization \\
A13 & $1.23 \pm 0.10$ & {B1, B2} & $20.45 \pm 0.05$, $20.16 \pm 0.05$ & $1.45 \pm 0.01$, $1.43 \pm 0.01$ & Ineffective Specialization \\
A15 & $2.77 \pm 0.15$ & {B2, B1} & $19.55 \pm 0.06$, $18.99 \pm 0.06$ & $1.23 \pm 0.00$, $1.20 \pm 0.00$ & Ineffective Consensus \\
A08 & $3.13 \pm 0.17$ & {B1, B0} & $20.19 \pm 0.05$, $19.43 \pm 0.06$ & $1.22 \pm 0.01$, $1.18 \pm 0.01$ & Ineffective Specialization \\
A12 & $7.91 \pm 0.20$ & {B1, B0} & $24.00 \pm 0.08$, $19.12 \pm 0.04$ & $1.63 \pm 0.01$, $1.40 \pm 0.00$ & Ineffective Specialization \\
A17 & $14.27 \pm 0.40$ & {B1, B2} & $23.91 \pm 0.10$, $19.20 \pm 0.05$ & $1.59 \pm 0.01$, $1.37 \pm 0.00$ & {Flawed Specialization (Vulnerability)} \\
A10 & $17.28 \pm 0.34$ & {B2, B1} & $22.78 \pm 0.07$, $20.59 \pm 0.08$ & $1.66 \pm 0.01$, $1.56 \pm 0.01$ & Ineffective Specialization \\
A18 & $28.61 \pm 0.34$ & {B1, B2} & $24.24 \pm 0.13$, $20.97 \pm 0.08$ & $2.08 \pm 0.01$, $1.93 \pm 0.01$ & {Flawed Specialization (Vulnerability)} \\ \bottomrule
\end{tabularx}
\end{adjustwidth}%

\end{table}

\vspace{-9pt}

\begin{table}[H]
\caption{{Descriptive} 
 statistics of the dominant contribution share (\%) stratified by operational archetype.}
\label{tab:dominant_share_by_archetype}
\begin{tabularx}{\textwidth}{lRRRR}
\toprule
\textbf{Identified Strategy} & \textbf{Mean} & \textbf{Std} & \textbf{Var} & \textbf{Count} \\ \midrule
Effective Consensus      & 18.76 & 0.63 & 0.39 & 3 \\
Effective Specialization & 23.18 & 2.00 & 4.02 & 3 \\
Flawed Specialization    & 23.81 & 2.51 & 6.29 & 3 \\
Ineffective Consensus    & 18.84 & 0.53 & 0.29 & 4 \\ \bottomrule
\end{tabularx}
\end{table}

\vspace{-9pt}

\begin{table}[H]
\caption{Summed Mean SHAP Values per Branch/Component with 95\% confidence intervals reported as $\text{value} \pm \text{CI}$.}
\label{tab:appendix_shap_sum}
\small
\begin{tabularx}{\textwidth}{lRRRRRR}
\toprule
\textbf{Attack} & \boldmath{$\Phi_{B0}$} & \boldmath{$\Phi_{B1}$} & \boldmath{$\Phi_{B2}$} & \boldmath{$\Phi_{B3}$} & \boldmath{$\Phi_{GAT-S}$} & \boldmath{$\Phi_{GAT-T}$} \\ 
\midrule
A07 & $1.79 \pm 0.01$ & $1.87 \pm 0.01$ & $1.95 \pm 0.00$ & $1.75 \pm 0.01$ & $1.50 \pm 0.01$ & $0.94 \pm 0.01$ \\
A08 & $1.35 \pm 0.00$ & $1.42 \pm 0.00$ & $1.37 \pm 0.01$ & $1.28 \pm 0.00$ & $0.74 \pm 0.01$ & $0.86 \pm 0.01$ \\
A09 & $2.92 \pm 0.01$ & $2.82 \pm 0.01$ & $3.35 \pm 0.01$ & $2.77 \pm 0.01$ & $1.40 \pm 0.01$ & $1.23 \pm 0.00$ \\
A10 & $1.60 \pm 0.01$ & $1.83 \pm 0.01$ & $1.96 \pm 0.01$ & $1.47 \pm 0.00$ & $1.05 \pm 0.01$ & $1.59 \pm 0.01$ \\

\bottomrule
\end{tabularx}
\end{table}

\begin{table}[H]\ContinuedFloat
\small
\caption{{\em Cont.}}
\begin{tabularx}{\textwidth}{lRRRRRR}
\toprule
\textbf{Attack} & \boldmath{$\Phi_{B0}$} & \boldmath{$\Phi_{B1}$} & \boldmath{$\Phi_{B2}$} & \boldmath{$\Phi_{B3}$} & \boldmath{$\Phi_{GAT-S}$} & \boldmath{$\Phi_{GAT-T}$} \\ 
\midrule

A11 & $1.72 \pm 0.01$ & $1.52 \pm 0.00$ & $1.74 \pm 0.01$ & $1.53 \pm 0.00$ & $1.47 \pm 0.01$ & $1.00 \pm 0.01$ \\
A12 & $1.59 \pm 0.01$ & $1.89 \pm 0.01$ & $1.62 \pm 0.00$ & $1.46 \pm 0.01$ & $1.16 \pm 0.01$ & $0.84 \pm 0.01$ \\
A13 & $1.58 \pm 0.00$ & $1.69 \pm 0.01$ & $1.70 \pm 0.01$ & $1.61 \pm 0.01$ & $1.23 \pm 0.01$ & $0.72 \pm 0.00$ \\
A14 & $2.10 \pm 0.01$ & $1.84 \pm 0.01$ & $2.28 \pm 0.01$ & $1.66 \pm 0.01$ & $1.63 \pm 0.01$ & $1.15 \pm 0.01$ \\
A15 & $1.38 \pm 0.01$ & $1.42 \pm 0.01$ & $1.49 \pm 0.01$ & $1.27 \pm 0.00$ & $1.07 \pm 0.01$ & $1.10 \pm 0.01$ \\
A16 & $1.47 \pm 0.00$ & $1.48 \pm 0.01$ & $1.50 \pm 0.00$ & $1.41 \pm 0.01$ & $1.22 \pm 0.01$ & $0.76 \pm 0.00$ \\
A17 & $1.61 \pm 0.01$ & $1.89 \pm 0.01$ & $1.65 \pm 0.01$ & $1.51 \pm 0.00$ & $1.15 \pm 0.01$ & $0.89 \pm 0.01$ \\
A18 & $2.60 \pm 0.01$ & $2.51 \pm 0.01$ & $2.39 \pm 0.01$ & $2.29 \pm 0.01$ & $1.38 \pm 0.01$ & $1.36 \pm 0.01$ \\
A19 & $1.54 \pm 0.01$ & $1.69 \pm 0.01$ & $1.76 \pm 0.00$ & $1.62 \pm 0.01$ & $1.38 \pm 0.01$ & $1.16 \pm 0.01$ \\ \bottomrule
\end{tabularx}

\end{table}
\vspace{-9pt}

\begin{table}[H]
\caption{Contribution Shares (\%) per Component after Softmax Normalization. Confidence intervals are provided in the format {value $\pm$ CI} 
 for each~value.}
\label{tab:appendix_softmax}
\small
\begin{tabularx}{\textwidth}{lRRRRrr}
\toprule
\textbf{Attack} & \textbf{B0 Share} & \textbf{B1 Share} & \textbf{B2 Share} & \textbf{B3 Share} & \textbf{GAT-S Share} & \textbf{GAT-T Share} \\ \midrule
A07 & $17.9 \pm 0.05$ & $18.8 \pm 0.07$ & $22.6 \pm 0.10$ & $18.1 \pm 0.07$ & $13.1 \pm 0.07$ & $9.4 \pm 0.03$ \\
A08 & $19.4 \pm 0.05$ & $20.2 \pm 0.05$ & $19.0 \pm 0.06$ & $18.2 \pm 0.04$ & $11.0 \pm 0.03$ & $12.1 \pm 0.04$ \\
A09 & $20.6 \pm 0.09$ & $21.9 \pm 0.13$ & $22.9 \pm 0.12$ & $20.1 \pm 0.11$ & $7.8 \pm 0.07$ & $6.8 \pm 0.07$ \\
A10 & $17.3 \pm 0.06$ & $20.6 \pm 0.07$ & $22.8 \pm 0.07$ & $15.3 \pm 0.05$ & $10.2 \pm 0.02$ & $13.9 \pm 0.08$ \\
A11 & $19.5 \pm 0.07$ & $18.0 \pm 0.07$ & $19.2 \pm 0.07$ & $18.2 \pm 0.05$ & $14.2 \pm 0.09$ & $11.0 \pm 0.04$ \\
A12 & $19.1 \pm 0.04$ & $24.0 \pm 0.08$ & $19.0 \pm 0.05$ & $17.4 \pm 0.05$ & $10.8 \pm 0.04$ & $9.7 \pm 0.03$ \\
A13 & $19.4 \pm 0.05$ & $20.4 \pm 0.05$ & $20.2 \pm 0.05$ & $18.8 \pm 0.06$ & $12.1 \pm 0.06$ & $9.1 \pm 0.03$ \\
A14 & $18.5 \pm 0.07$ & $16.4 \pm 0.06$ & $26.2 \pm 0.10$ & $16.7 \pm 0.07$ & $12.0 \pm 0.08$ & $10.2 \pm 0.03$ \\
A15 & $18.6 \pm 0.05$ & $19.0 \pm 0.06$ & $19.6 \pm 0.05$ & $17.2 \pm 0.04$ & $12.6 \pm 0.04$ & $13.1 \pm 0.04$ \\
A16 & $19.0 \pm 0.05$ & $19.2 \pm 0.05$ & $19.8 \pm 0.05$ & $18.5 \pm 0.04$ & $13.3 \pm 0.05$ & $10.2 \pm 0.03$ \\
A17 & $17.8 \pm 0.05$ & $23.9 \pm 0.10$ & $19.2 \pm 0.05$ & $17.5 \pm 0.06$ & $12.0 \pm 0.03$ & $9.5 \pm 0.03$ \\
A18 & $19.4 \pm 0.15$ & $24.2 \pm 0.12$ & $21.0 \pm 0.09$ & $18.5 \pm 0.08$ & $8.3 \pm 0.05$ & $8.6 \pm 0.05$ \\
A19 & $17.5 \pm 0.05$ & $20.1 \pm 0.06$ & $20.0 \pm 0.08$ & $17.9 \pm 0.08$ & $12.2 \pm 0.07$ & $12.2 \pm 0.05$ \\ \bottomrule
\end{tabularx}
\end{table}

\subsection{Justification of the Analysis~Method}
The proposed methodology integrates spectral theory with game-theoretic attribution to provide a robust interpretation framework. We analyze the eigenvalue spectrum of covariance matrices because features derived from covariance functionals are statistically stable and converge at a rate of \( O(N^{-1/2}) \) \cite{sihag2023covarianceneuralnetworks}. This robustness is crucial when analyzing deep layers where activations can be noisy. Furthermore, recent studies indicate that adversarial perturbations often manifest as shifts toward the null eigenspace of empirical kernels~\cite{harzli2025adversarial}, suggesting that spectral features are particularly sensitive to the artifacts introduced by spoofing attacks. By~coupling these robust features with SHAP~\cite{Lundberg2017unified}, which guarantees the fair distribution of prediction credit among features, we ensure that our attribution of ``importance'' to specific AASIST3~\cite{Borodin2024aasist3} branches is mathematically principled and not an artifact of the visualization~method.

{It is important to clarify the epistemological scope and methodological boundaries of the proposed interpretability framework. Our spectral-SHAP approach operates at a meta-level, establishing correlational structures between empirically observable activation patterns and downstream performance outcomes, rather than providing direct causal attributions of AASIST3's internal computational mechanisms. The~SHAP values derived from the CatBoost meta-classifier quantify statistical associations between spectral signatures from intermediate layers and classification success on specific attack types, capturing how the architecture behaves under different input distributions, not why particular feature representations emerged during training. These interpretations reflect emergent operational strategies---consensus, specialization, or~internal conflict---as patterns in the attribution landscape, without~attributing mechanistic causality to individual neurons, attention heads, or~graph convolutions. Consequently, our findings are conditioned on the specific AASIST3 instantiation we analyzed; different training procedures or datasets may yield different operational archetypes even with identical architecture, because~learned feature representations would differ. This methodological position is analogous to behavioral modeling in complex systems: we characterize what the model does in its current configuration, not what the architecture is fundamentally capable of doing across all \mbox{possible parameterizations}.}

\section{Results}
\label{sec:results}

In this section, we correlate the internal operational strategies of AASIST3~\cite{Borodin2024aasist3} with its objective performance. Performance is evaluated using the Equal Error Rate (EER), defined as the threshold point where the False Acceptance Rate (FAR) equals the False Rejection Rate (FRR), Equation~(\ref{eq:eer}).
\begin{equation}\label{eq:eer}
    \text{EER} = \text{FAR}(\theta) = \text{FRR}(\theta)
\end{equation}
where \(\theta\) is the decision threshold. A~lower EER indicates better detection performance on a specific attack type from the ASVspoof dataset~\cite{wang2020asvspoof2019largescalepublic}.

\subsection{Eigenvalue Count~Ablation}\label{sec:eigenvalues_ablations}

\begin{figure}[H]
    \includegraphics[width=.95\linewidth]{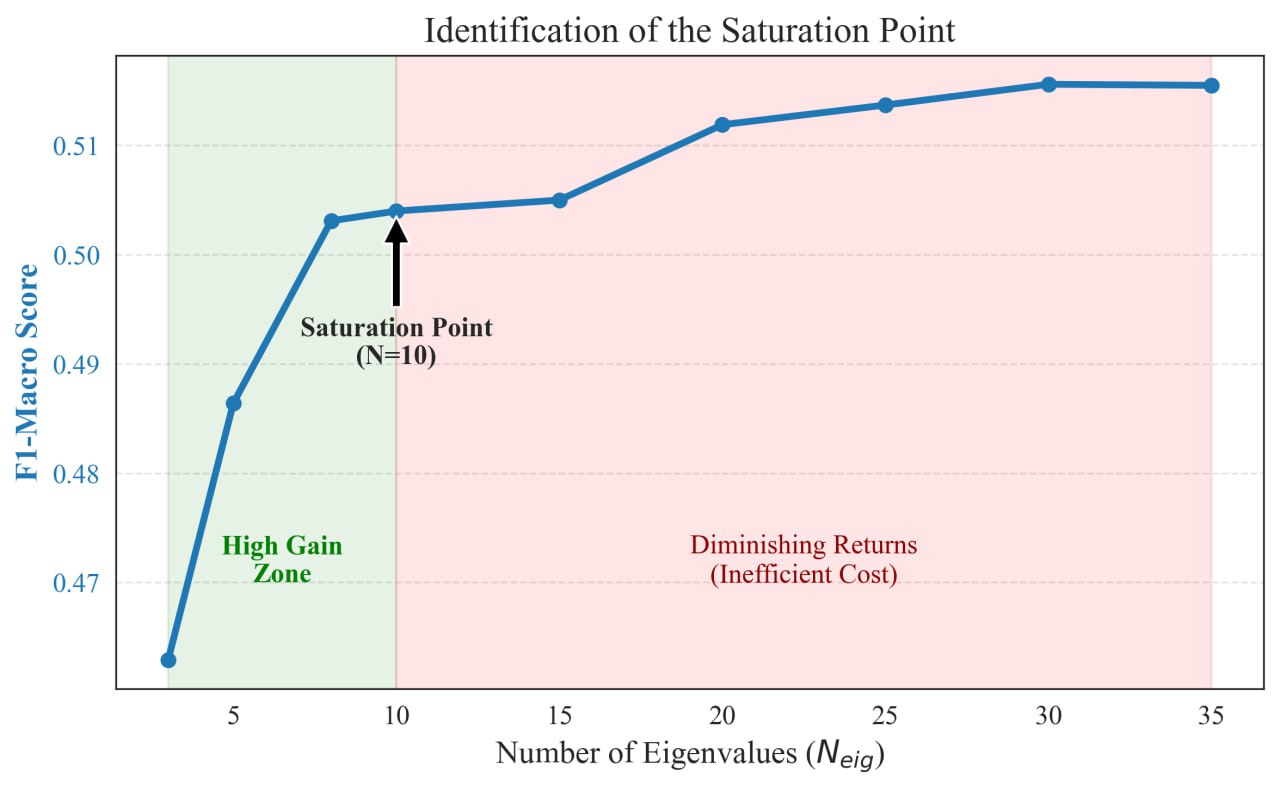}
    \caption{Identification of the saturation point for the number of retained eigenvalues $N_eig$.}
    \label{fig:eig_perf_vs_number}
\end{figure}

\begin{figure}[H]
    \includegraphics[width=.95\linewidth]{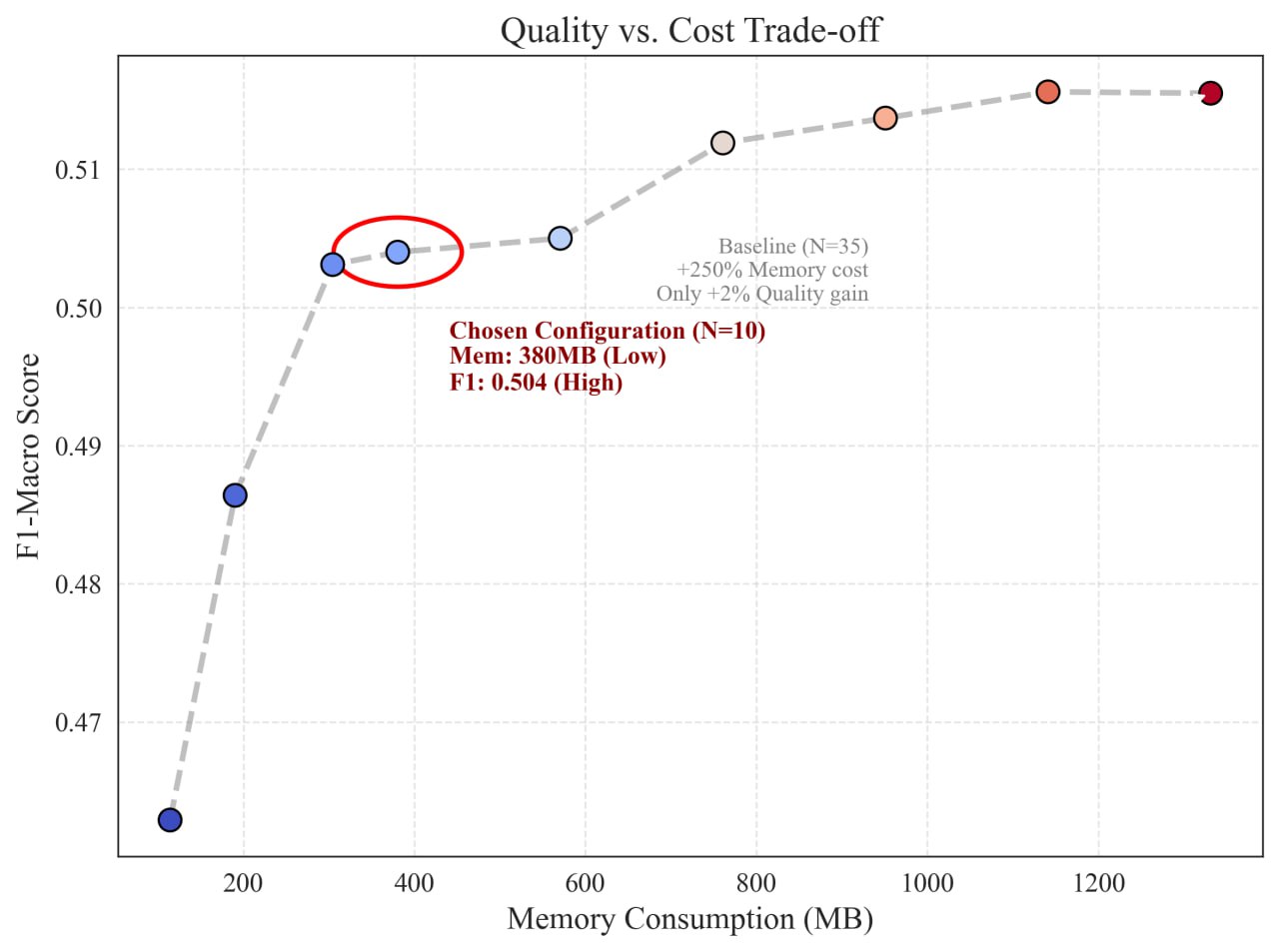}
    \caption{{Quality–cost} 
 trade-off between F1-Macro score and memory consumption across different eigenvalue~counts.}
    \label{fig:eig_perf_vs_mem}
\end{figure}

\begin{figure}[H]
    \includegraphics[width=.95\linewidth]{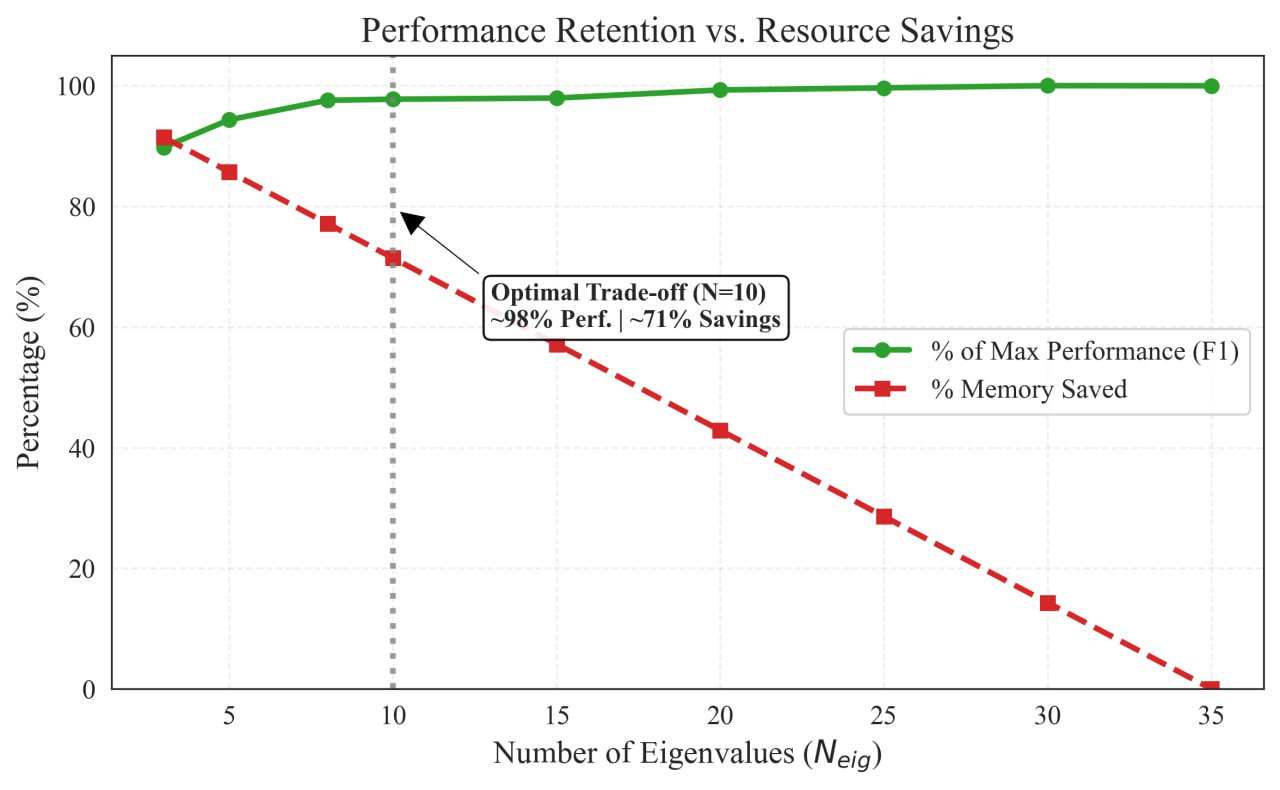}
    \caption{{Performance} 
 retention and memory savings as a function of the number of retained eigenvalues.}
    \label{fig:eig_perf_in_prov_vs_num}
\end{figure}

{We evaluate the sensitivity of the proposed spectral signature to the number of retained eigenvalues $N_{eig}$ by performing an ablation study that jointly considers predictive performance and resource usage (Figures~\ref{fig:eig_perf_vs_number}--\ref{fig:eig_perf_in_prov_vs_num}). The~results indicate a clear saturation behavior: increasing $N_{eig}$ yields substantial gains in the low-$N_{eig}$  regime, while improvements become marginal beyond $N_{eig}=10$, where the curve enters a diminishing-returns region. In~addition, the~performance–memory trade-off demonstrates that $N_{eig}=10$ provides an effective operating point, retaining approximately 98\% of the maximum F1-Macro score while achieving about 71\% memory savings compared with larger configurations. Accordingly, we set $N_{eig}=10$ in all subsequent experiments as a principled compromise between accuracy and computational cost.}

\vspace{-3pt}

{Figure~\ref{fig:eig_perf_vs_number} illustrates the relationship between the number of eigenvalues $N_{eig}$  selected for the spectral signature and the resulting F1-Macro score. The~plot is divided into a high-gain zone, where the performance increases sharply as $N_{eig}$  moves from 2 toward 10, and~a subsequent region of diminishing returns for values exceeding 10. A~saturation point is explicitly identified at $N_{eig}=10$, marking the transition from rapid performance growth to a plateau where additional computational costs result in marginal quality gains.
}

\vspace{-3pt}

\vspace{-9pt}

{Figure~\ref{fig:eig_perf_vs_mem} presents a scatter plot of F1-Macro score versus memory consumption (MB) for configurations obtained by varying the number of retained eigenvalues $N_{eig}$. Each point corresponds to a specific $N_{eig}$ setting, and~the dashed line connects configurations in increasing computational cost, visualizing how model quality changes with memory usage. The~figure also annotates a selected operating point at $N_{eig}=10$ (shown together with its reported memory value and F1-Macro score) and includes a reference annotation for a larger baseline configuration (e.g., $N_{eig}=35$).}

{Figure~\ref{fig:eig_perf_in_prov_vs_num} depicts two curves summarizing the effect of varying the number of retained eigenvalues $N_{eig}$  on both predictive performance and resource usage. The~green curve reports the F1 score expressed as a percentage of the maximum observed performance, while the red curve reports the percentage of memory saved relative to the highest-cost configuration, both plotted against $N_{eig}$. The~figure highlights a reference point at $N_{eig}=10$ using a vertical guideline and an annotation box that reports the corresponding performance-retention and memory-savings values.}

\subsection{Penalty Function~Ablation}

\begin{figure}[H]
    \includegraphics[width=.77\linewidth]{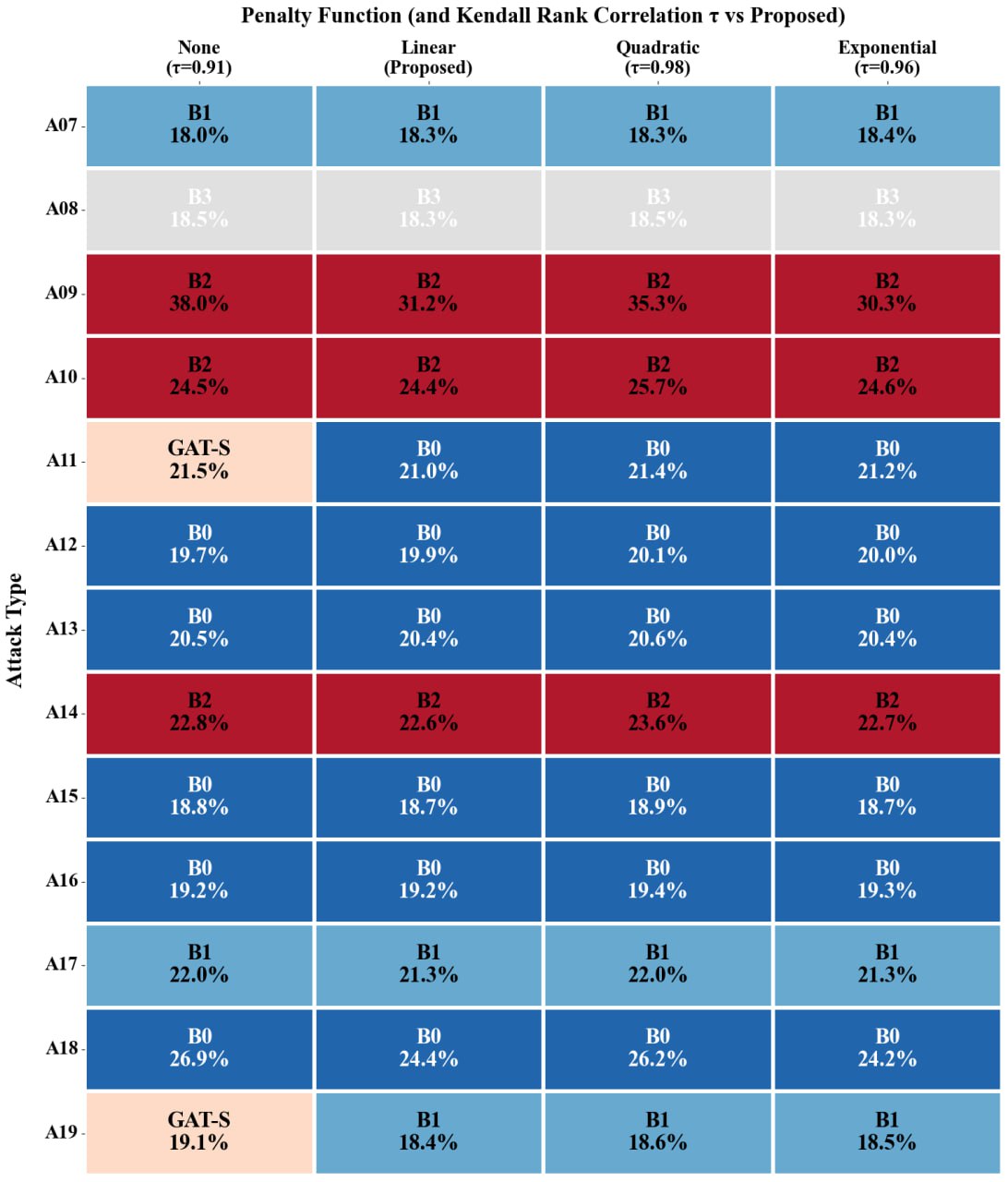}
    \caption{{Ablation} 
 study of the penalty function. Colors represent the identity of the dominant branch (Blue shades: Branches B0/B1; Red shades: Branches B2
; Grey shades: B3; Apricot shades: Graph attention modeuls).
Consistency of color across a row indicates that the identification of the dominant branch is robust to the choice of penalty function. $\tau$ denotes the Kendall rank correlation coefficient comparing the branch ranking of each method against our proposed `Linear' penalty.}
    \label{fig:penalty_function_ablation}
\end{figure}

{Figure~\ref{fig:penalty_function_ablation} presents the results of the penalty function ablation study. The~heatmap visualizes the dominant branch identified for each attack type (A07–A19) under the \mbox{four different} penalty regimes. The~consistency of the color coding across each row confirms that the core strategic findings are largely invariant to the specific form of the penalty. For~instance, Attacks A09, A14, and~A10 consistently identify Branch B2 (red shades) as dominant, while A12, A13, A15, A16, and~A18 consistently point to Branch B0 (blue shades).}

\vspace{-3pt}

{Quantitatively, the~proposed linear penalty achieves a high agreement with the stricter quadratic ($\tau=0.98$) and exponential ($\tau=0.96$) formulations, indicating that it provides a balanced penalization that effectively filters out unstable branches without suppressing valid but slightly noisy signals. The~``None'' baseline shows a slightly lower correlation ($\tau=0.91$), highlighting that some form of variance penalization is indeed necessary to resolve ambiguities in attacks like A11 and A19, where raw SHAP values alone may lead to inconsistent attributions (e.g., flipping between GAT-S and B0/B1).
}

\subsection{Overview of Model Performance and Internal~Strategies}
Table~\ref{tab:main_analysis} synthesizes the key metrics for each attack. The~attacks are sorted by their EER to highlight the relationship between difficulty and model strategy. The~``Dominant Share'' column is derived using the Softmax-normalized confidence metric defined in \mbox{Equation~(\ref{eq:dominant_shate})}, while the ``Confidence score'' column reports the raw $C_b$ values from Equation~(\ref{eq:confidence_score}).  This score explicitly accounts for internal stability: lower values indicate higher internal variance or conflict among the components of a branch, providing quantitative evidence that certain attacks induce disagreement within the model's decision-making process even when contribution shares appear balanced. The ``Identified Strategy'' column classifies the model's overall behavior according to the operational archetypes defined in Section~\ref{sec:materials}.

\begin{figure}[H]
    \includegraphics[width=.99\linewidth]{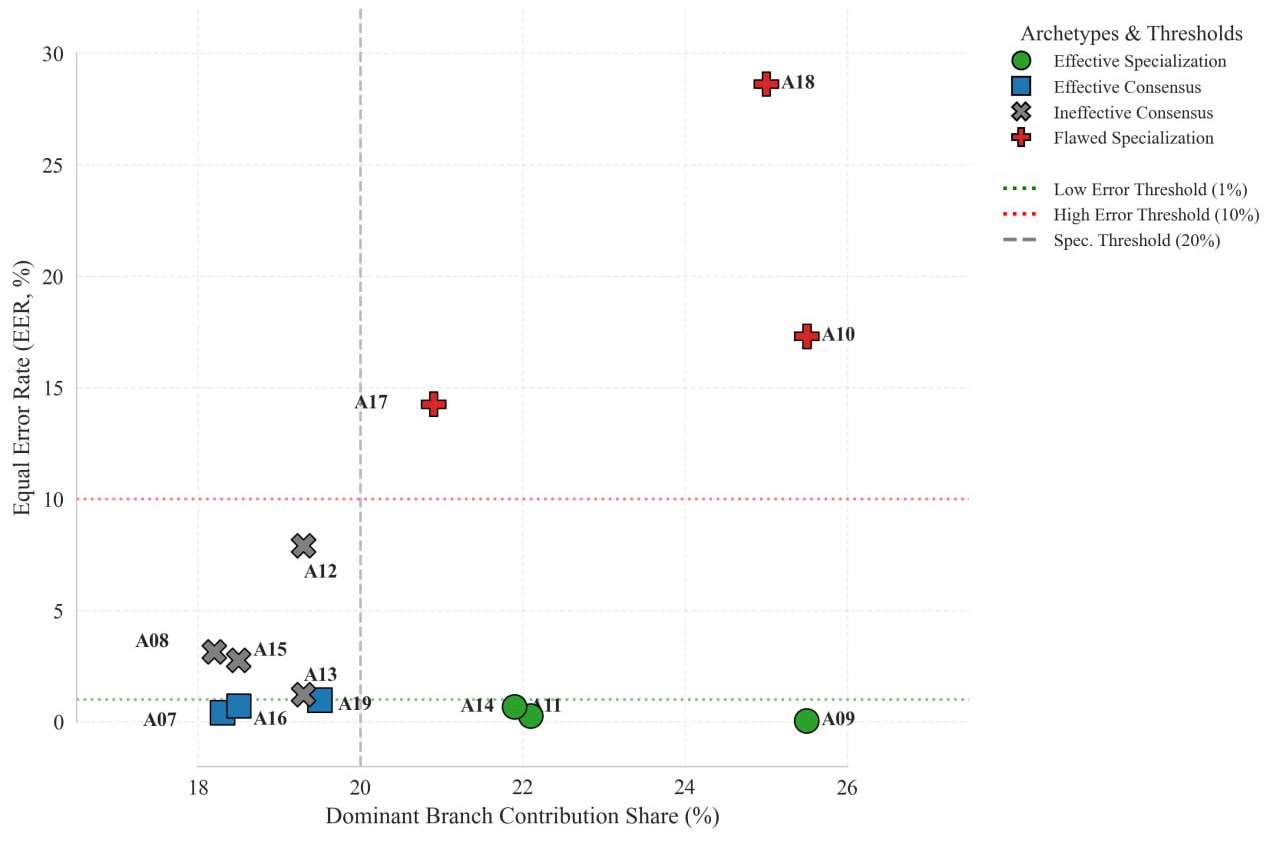}
    \caption{{Internal} 
 Strategy Matrix: categorization of attacks based on the empirical relationship between Equal Error Rate (EER) and Dominant Branch Contribution~Share.}
    \label{fig:internal_strategy_matrix}
\end{figure}

{Figure~\ref{fig:internal_strategy_matrix} visualizes the operational strategies of the AASIST3 model by mapping each attack into a 2D space defined by the EER and the contribution share of the dominant branch. This visualization provides the empirical rationale for our categorization thresholds: the 1\% EER boundary (green dotted line) cleanly separates reliable performance from failure modes, while the 20\% contribution threshold (vertical dashed line) effectively distinguishes distributed consensus from specialized reliance on a single component. By~segmenting the performance landscape into these four quadrants—ranging from high-accuracy ``Effective Specialization'' (bottom right) to the high-error failure mode of ``Flawed Specialization'' (top right)—the figure confirms that the chosen cutoffs are not arbitrary, but~rather reflect natural clusters in the model’s behavior under varying attack conditions.}

\vspace{-3pt}
\unskip

\subsection{Correlation and Variance Analysis of Operational~Archetypes}

To quantitatively assess the relationship between internal strategy and empirical performance, we conducted a correlation analysis between the dominant contribution share and the Equal Error Rate (EER) across all~attacks.

We computed both Pearson's linear correlation and Spearman's rank correlation between the dominant Softmax-normalized share and the corresponding EER, Equation~(\ref{eq:correlations}).
\begin{equation}\label{eq:correlations}
    r_{\text{Pearson}} = 0.537,\qquad
\rho_{\text{Spearman}} = 0.077.
\end{equation}

The moderate positive Pearson correlation indicates that, on~average, higher dominant shares tend to co-occur with higher EER, while the near-zero Spearman coefficient reveals that this effect is not monotonic when attacks are ranked by difficulty. In~other words, a~large dominant share does not systematically guarantee either low or high error; its impact is modulated by the broader operational archetype (e.g., whether the dominant branch acts as a correct expert or as a confidently wrong failure mode). This aligns with the qualitative findings in Figure~\ref{fig:internal_strategy_matrix}, where both Effective and Flawed Specialization occupy regions with comparable dominant shares but drastically different EER~levels.

We further analyzed how the dominant share behaves within each of the four operational archetypes by grouping attacks according to their identified strategy and computing basic descriptive statistics of the dominant~share.

The variance analysis reveals a clear separation between consensus- and specialization-based regimes. Both Effective and Ineffective Consensus exhibit relatively low dominant shares ($\approx$18.8\%) with very small variance, reflecting a balanced distribution of responsibility across branches regardless of whether the resulting decision is accurate. In~contrast, both specialization regimes show substantially higher dominant shares ($\approx$23\text{--}24\%) and larger variance, indicating that the model increasingly concentrates its trust in a single branch when it commits to a specialized strategy. Flawed Specialization displays the highest variance (\(\mathrm{var} = 6.29\)), suggesting that failure cases are associated not only with strong reliance on one branch but also with increased instability in how strongly this dominance manifests across attacks. This supports our interpretation that specialization is a double-edged mechanism: it underpins the best-performing attacks but also amplifies vulnerability when the model specializes on the wrong~expert.

\subsection{Detailed Per-Attack~Analysis}
The following subsections examine the 13 attacks individually. We analyze the SHAP~\cite{Lundberg2017unified} distribution maps and decision plots to validate the strategies identified in Table~\ref{tab:main_analysis}.

In SHAP distribution plots, each bar corresponds to a model component (for example, one of the branches B0–B3, a~pooling layer, or~a global attention module), and~its length encodes the average magnitude of that component’s SHAP value across many samples, so longer bars indicate components that have a stronger overall influence on the prediction. The~sign of the value reflects the direction of the effect relative to the target class: in this study, positive values push the model toward the spoof hypothesis and negative values toward bonafide, so a plot with uniformly large positive bars for one component means the model is consistently relying on that component as an “expert” for detecting artifacts of a particular attack, whereas mixed positive and negative contributions across components reveal internal disagreement or~confusion. 

Decision plots complement this by showing how these contributions accumulate along the model’s output axis: starting from a baseline score (often near zero), the~line traces the step‑by‑step change in the prediction as each component is added in order of importance, with~upward segments indicating components that increase spoof confidence and downward segments indicating components that suppress it. When the trajectory is smooth and dominated by a small set of components, the~plot visualizes effective specialization, where the decision is largely driven by one branch or module; when the line oscillates with many small conflicting steps, it reveals ineffective consensus, where no component provides a clear signal and the final prediction is the result of many weak, partially cancelling~influences.

\subsection{Interpretation Framework for SHAP~Analysis}

{To avoid redundancy, we establish a unified framework for interpreting the visualization results presented in the subsequent figures. The~analysis relies on two primary visualization types.}

{SHAP Distribution Maps (Violin Plots): These plots contrast the model's response to attack samples (red distributions, positive values pushing toward ``spoof'') versus bona fide samples (blue distributions, negative values pushing toward ``genuine''). A~clear vertical separation implies strong discriminative power, whereas overlapping or diffuse distributions indicate weak feature extraction.}

{Decision Plots: These visualize the cumulative decision path. A~steep, monotonic trajectory driven by one or two dominant components indicates Specialization. Conversely, a~chaotic, oscillating trajectory involving many small, conflicting steps signifies Ineffective Consensus, where the model lacks a clear signal and relies on the aggregation of \mbox{weak features}.}

\subsubsection{Attacks A07 and A08 (Consensus Strategies)}

\begin{figure}[H]
    \includegraphics[width=.99\linewidth]{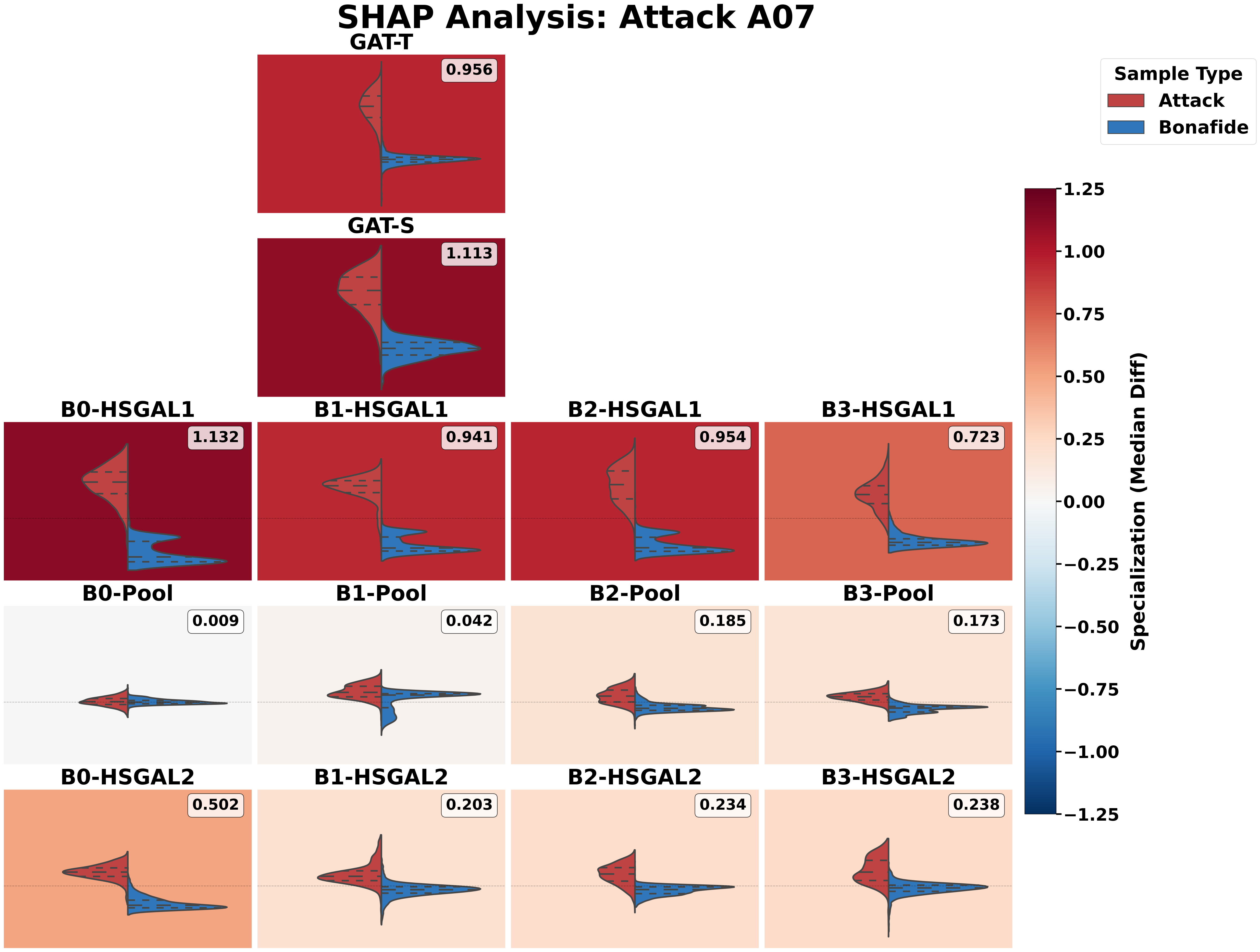}
    \caption{{SHAP} 
 Distribution for Attack A07. Red and blue distributions represent SHAP values for attack (positive, spoof-indicating) and bonafide (negative, genuine-indicating) samples, respectively.}
    \label{fig:shap_a07}
\end{figure}

\begin{figure}[H]
    \begin{minipage}{0.48\textwidth}
        \includegraphics[width=\linewidth]{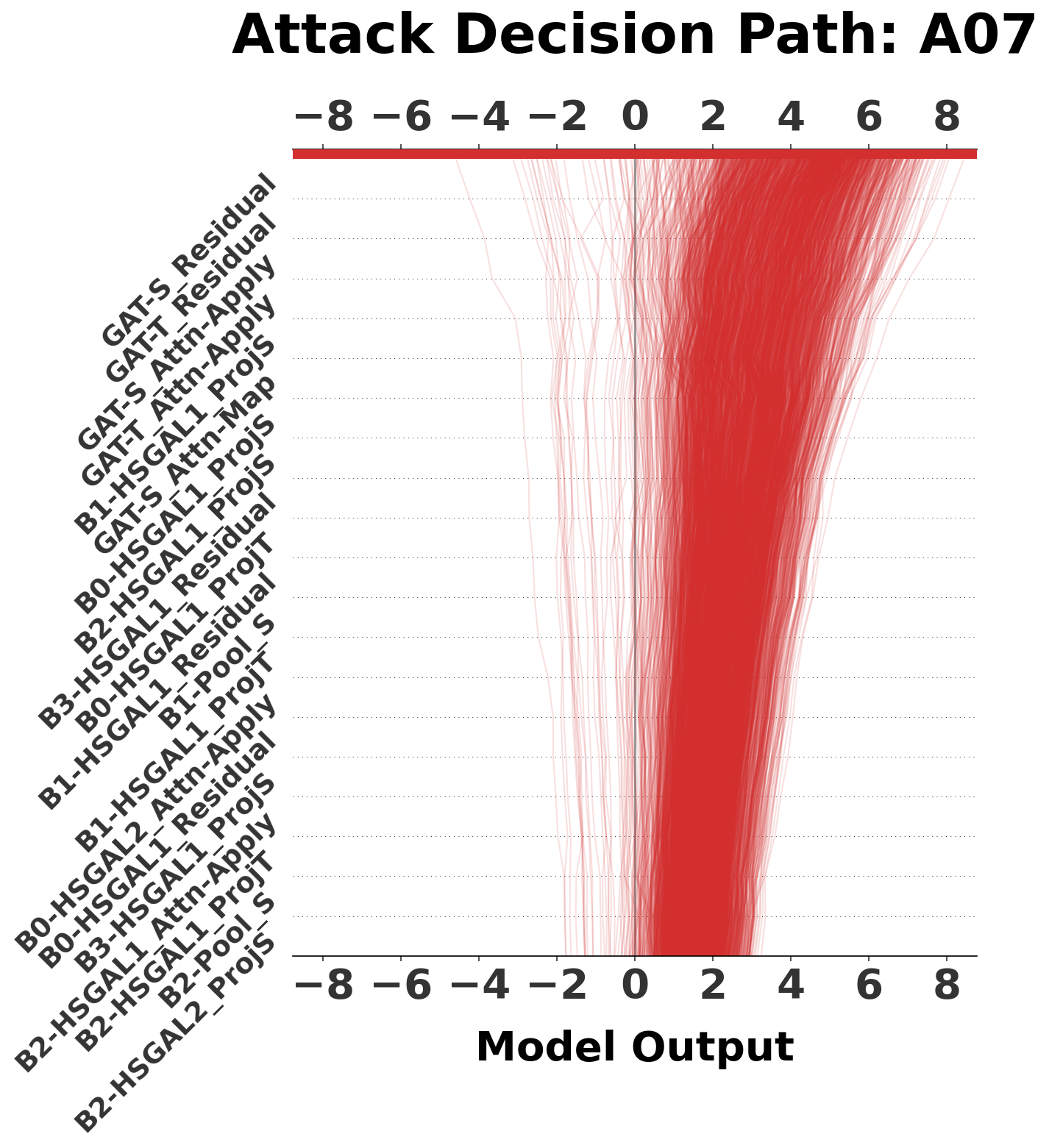}
        \caption*{\centering A07 Decision Plot}
    \end{minipage}%
   ~~~
    \begin{minipage}{0.48\textwidth}
        \includegraphics[width=\linewidth]{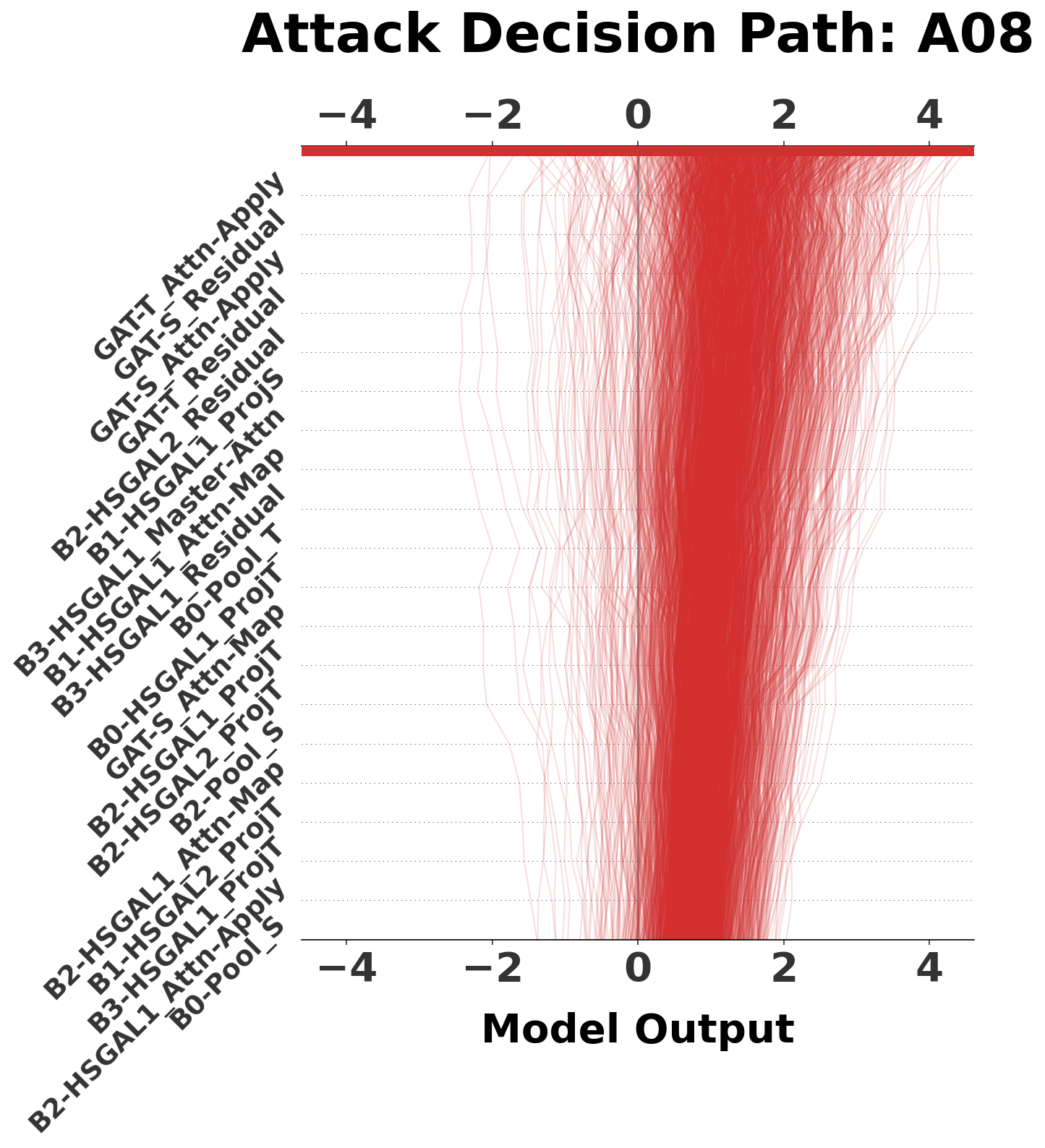}
        \caption*{\centering A08 Decision Plot}
    \end{minipage}
    \caption{Comparative Decision Plots for Attacks A07 (\textbf{Left}) and A08 (\textbf{Right}).}
    \label{fig:dec_a07_a08}
\end{figure}

\begin{figure}[H]
    \includegraphics[width=.99\linewidth]{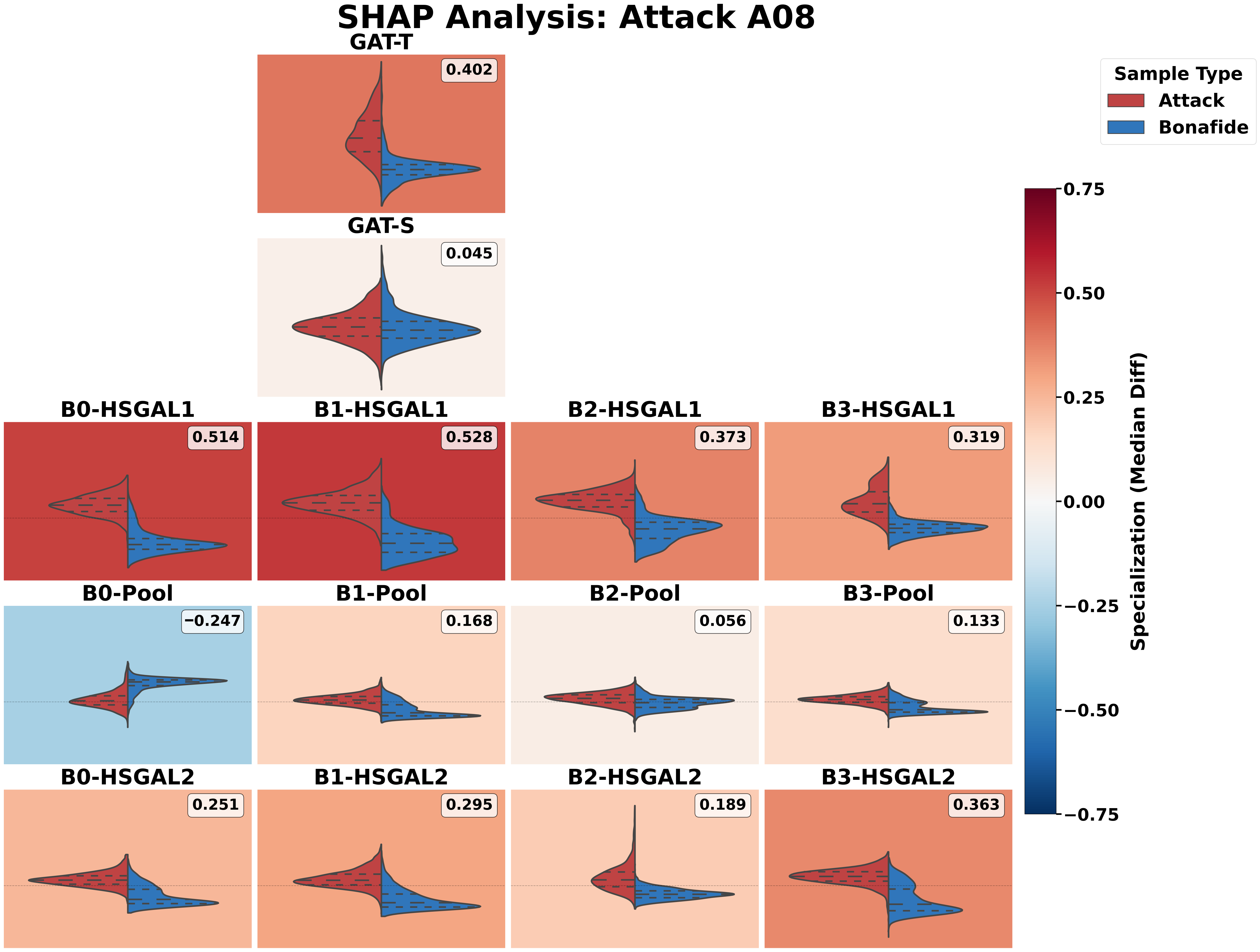}
    \caption{{SHAP} 
 Distribution for Attack A08. Red and blue distributions represent SHAP values for attack (positive, spoof-indicating) and bonafide (negative, genuine-indicating) samples, respectively.}
    \label{fig:shap_a08}
\end{figure}

{Attack A07 is handled via Effective Consensus, characterized by a moderate Confidence Score of $C_b=0.67$ and balanced contribution shares (Table~\ref{tab:main_analysis}). As~shown in Figure~\ref{fig:shap_a07}, virtually all components exhibit strong, positive SHAP contributions with distinct separation between attack and bona fide distributions. 

The~decision plot (Figure~\ref{fig:dec_a07_a08}, left) confirms this collaborative strategy, showing a smooth accumulation of evidence across both spectral and temporal branches.

In~contrast, Attack A08 triggers an Ineffective Consensus. While contribution shares remain balanced, the~Confidence Score drops significantly to $C_b=0.33$, the~lowest among all attacks (Table~\ref{tab:main_analysis}). The~SHAP distributions (Figure~\ref{fig:shap_a08}) are compressed near zero with notable overlap, and~the decision plot (Figure~\ref{fig:dec_a07_a08}, right) reveals a jagged trajectory. This visual instability is quantitatively captured by the low $C_b$ score, reflecting the model's struggle to find a robust artifact in any single domain.}

\subsubsection{Attacks A09 and A10 (Specialization vs. Complexity)}

\begin{figure}[H]
    \includegraphics[width=.99\linewidth]{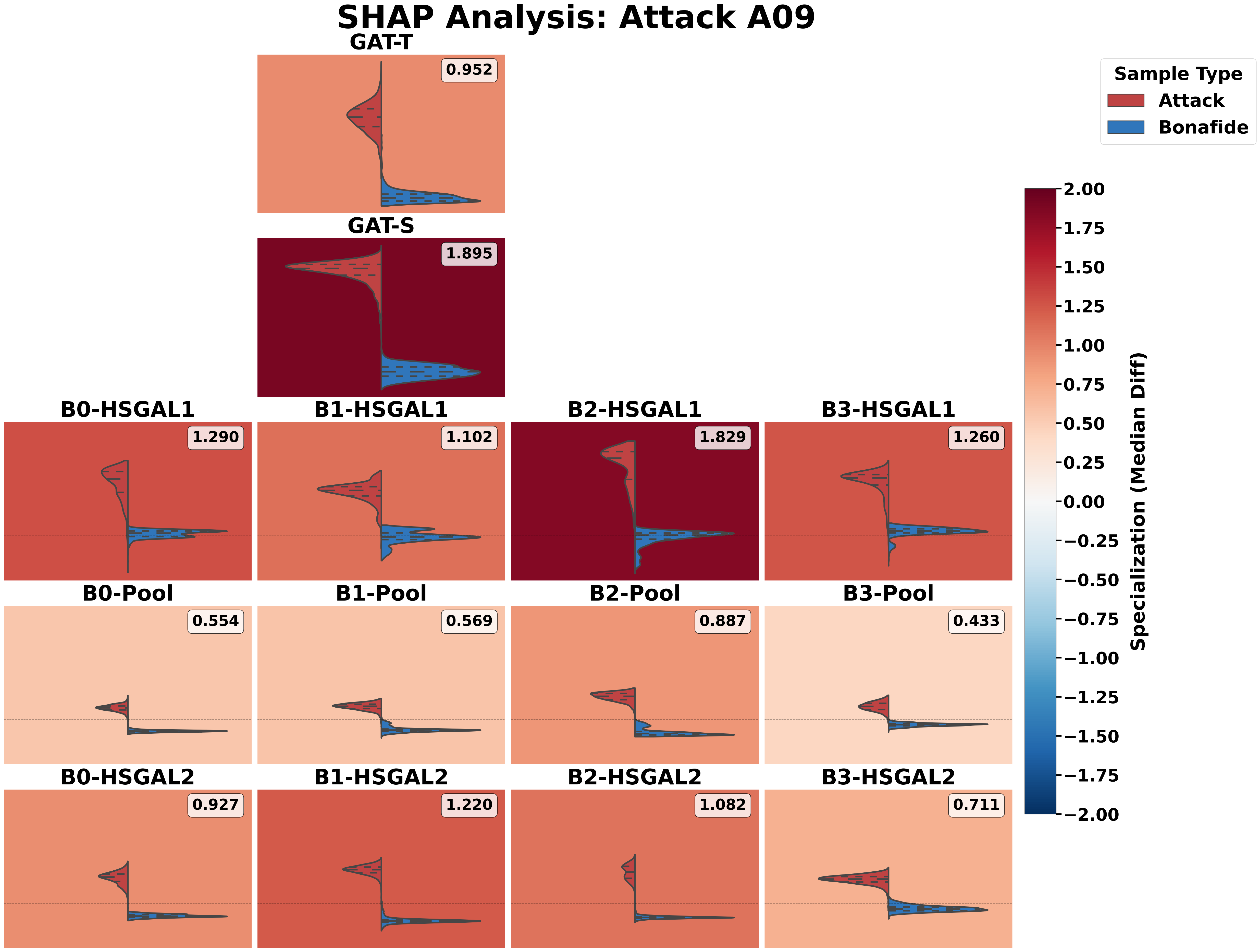}
    \caption{SHAP Distribution for Attack A09. Red and blue distributions represent SHAP values for attack (positive, spoof-indicating) and bonafide (negative, genuine-indicating) samples, respectively.}
    \label{fig:shap_a09}
\end{figure}

\begin{figure}[H]
    \includegraphics[width=.99\linewidth]{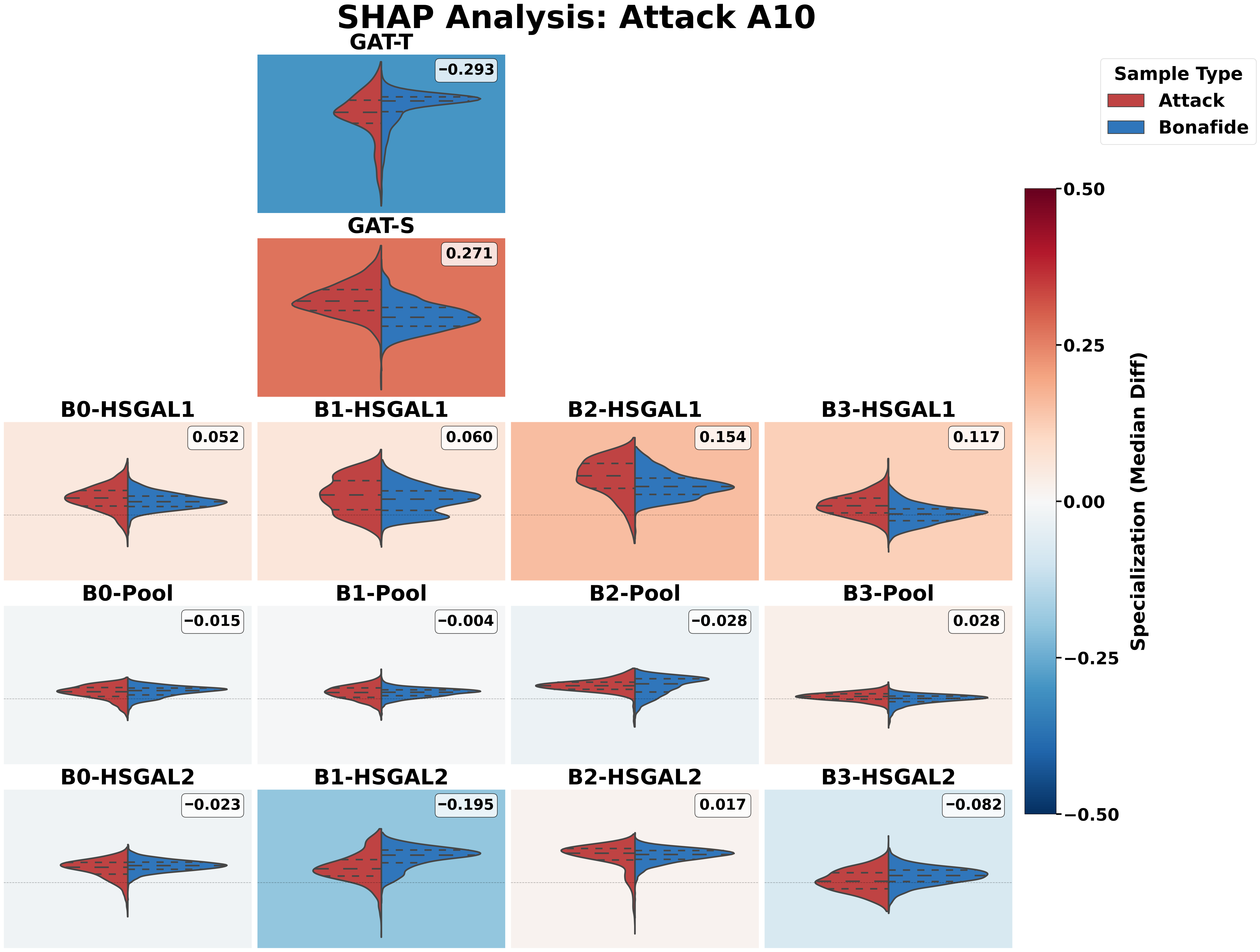}
    \caption{{SHAP} 
 Distribution for Attack A10. Red and blue distributions represent SHAP values for attack (positive, spoof-indicating) and bonafide (negative, genuine-indicating) samples, respectively.}
    \label{fig:shap_a10}
\end{figure}

\begin{figure}[H]
    \begin{minipage}{0.48\textwidth}
        \includegraphics[width=\linewidth]{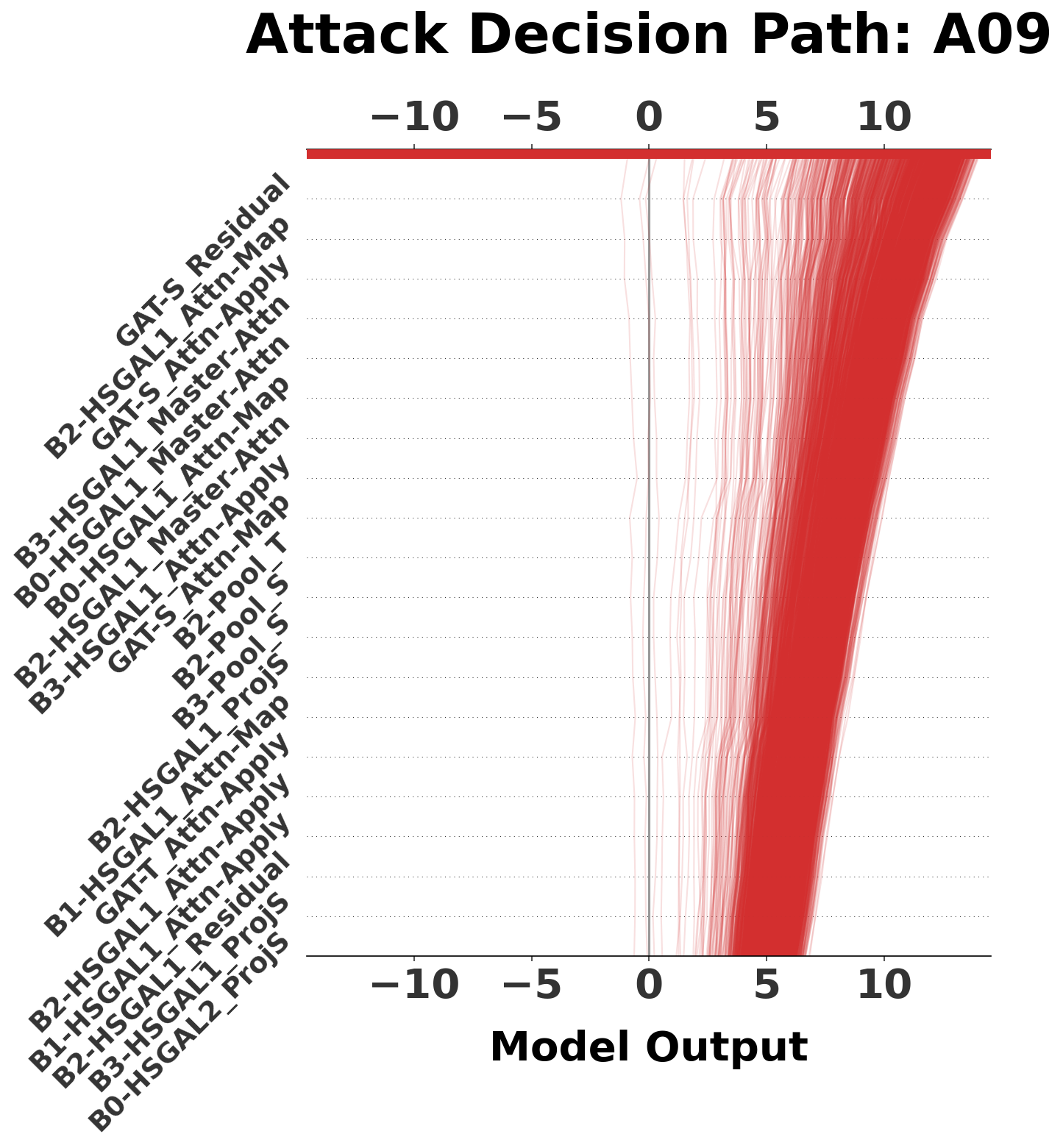}
        \caption*{\centering A09 Decision Plot}
    \end{minipage}%
   ~~~
    \begin{minipage}{0.47\textwidth}
        \includegraphics[width=\linewidth]{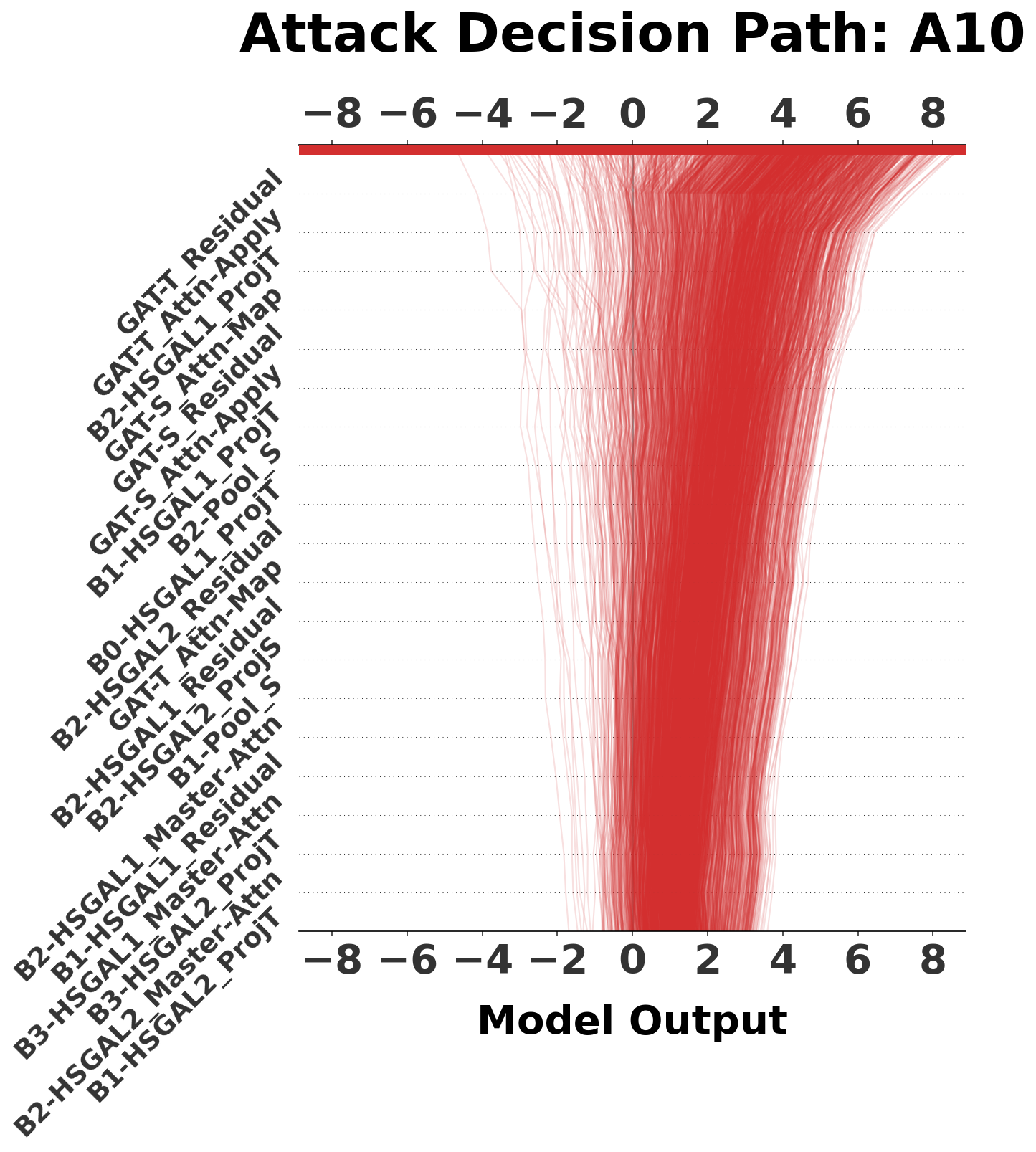}
        \caption*{\centering A10 Decision Plot}
    \end{minipage}
    \caption{Comparative Decision Plots for Attacks A09 (\textbf{Left}) and A10 (\textbf{Right}).}
    \label{fig:dec_a09_a10}
\end{figure}

{Attack A09 is a trivial case of Effective Specialization, quantitatively validated by the highest Confidence Score of $C_b=1.56$ and a dominant contribution share of 25.5\% from Branch B2 (Table~\ref{tab:main_analysis}). The~model delegates detection almost exclusively to B2, which shows an overwhelmingly large positive SHAP signal ({Figure}
~\ref{fig:shap_a09}). Conversely, Attack A10 (Figure~\ref{fig:shap_a10}) represents a failure case of Ineffective Consensus (reclassified from specialization due to high conflict). Although~the model attempts to prioritize GAT-T ($25.5\%$) and B2 ($21.9\%$), the~decision plot (Figure~\ref{fig:dec_a09_a10}, right) exposes extreme internal conflict, with~large opposing contributions canceling each other out. This internal disagreement is measurable: despite high raw contribution shares, the~Confidence Score is suppressed to $0.93$, confirming that while the model ``selects'' experts, their predictions are  unstable.}

\vspace{-6pt}

\subsubsection{Attacks A11 and A12 (Spectral vs. Mixed)}

\begin{figure}[H]
    \includegraphics[width=.98\linewidth]{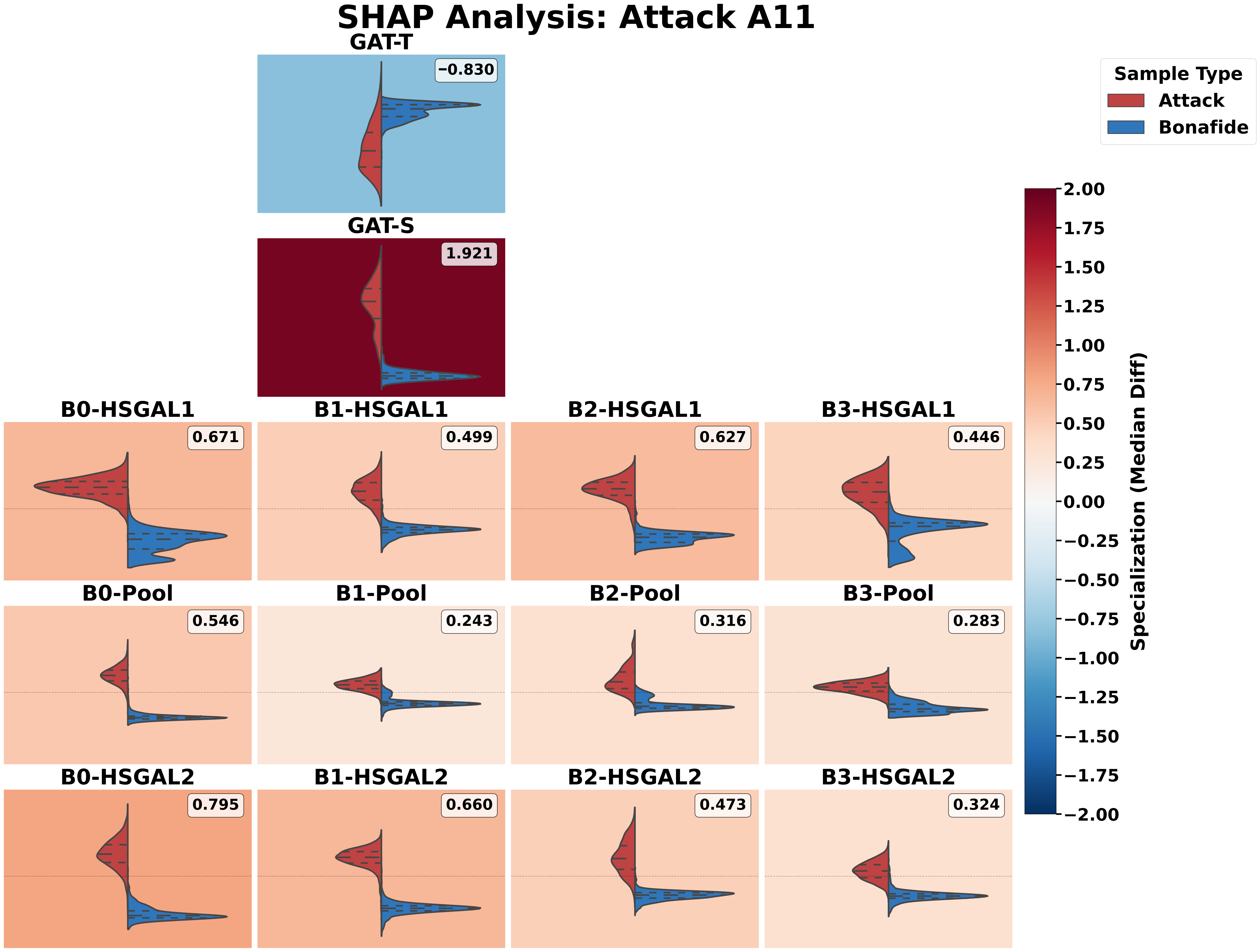}
    \caption{{SHAP} 
 Distribution for Attack A11. Red and blue distributions represent SHAP values for attack (positive, spoof-indicating) and bonafide (negative, genuine-indicating) samples, respectively.}
    \label{fig:shap_a11}
\end{figure}

\begin{figure}[H]
    \includegraphics[width=.98\linewidth]{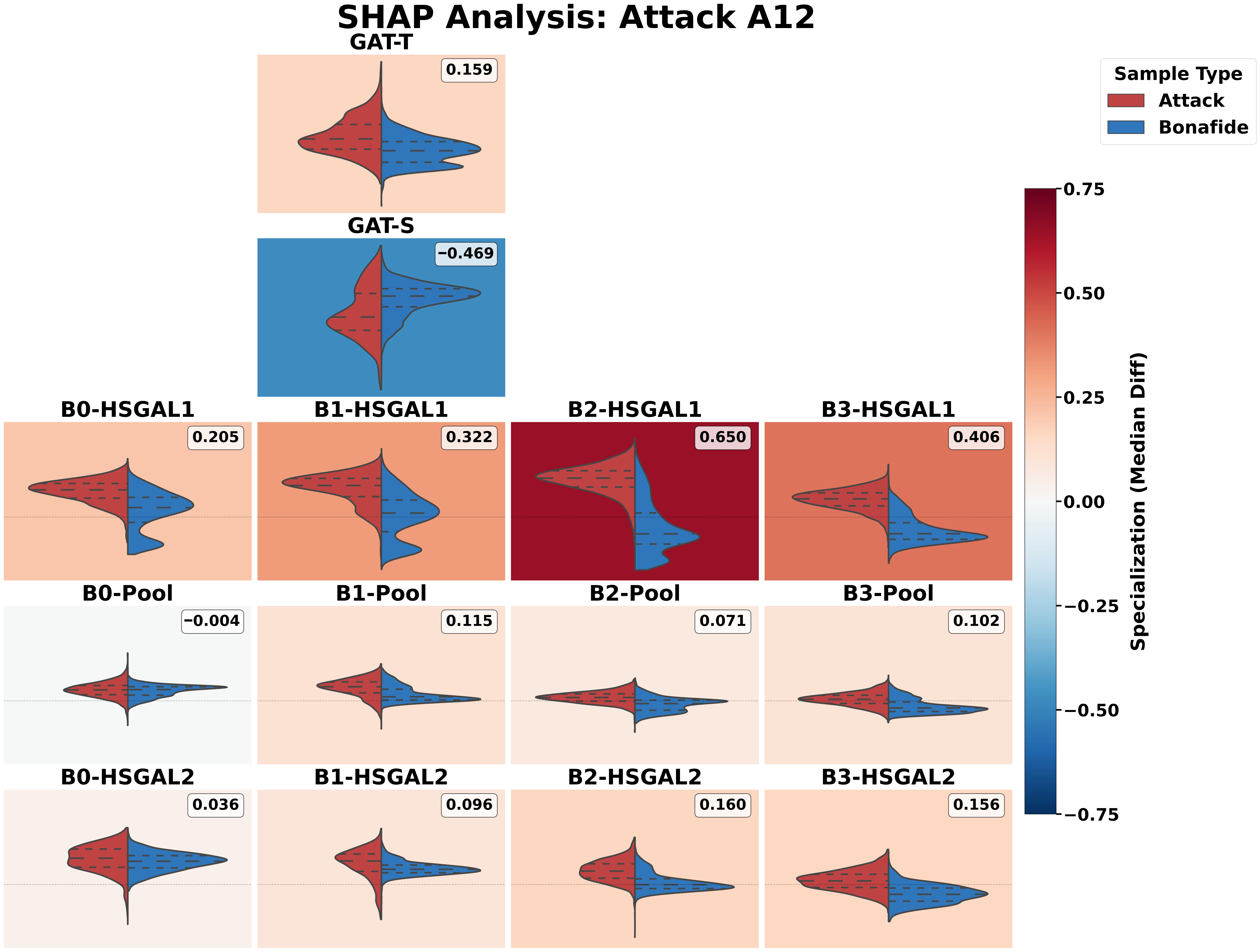}
    \caption{SHAP Distribution for Attack A12. Red and blue distributions represent SHAP values for attack (positive, spoof-indicating) and bonafide (negative, genuine-indicating) samples, respectively.}
    \label{fig:shap_a12}
\end{figure}

\begin{figure}[H]
    \begin{minipage}{0.485\textwidth}
        \includegraphics[width=\linewidth]{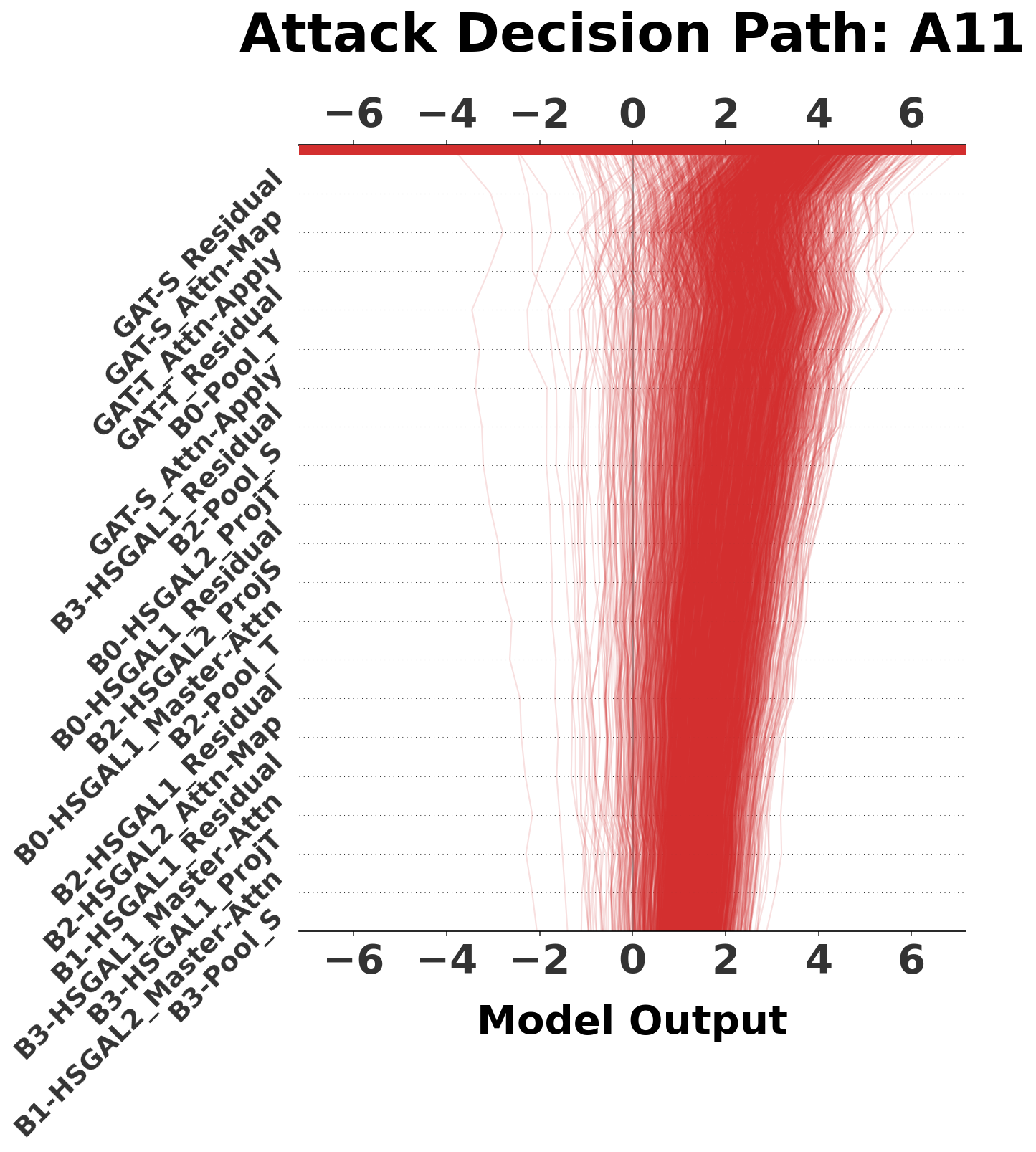}
        \caption*{\centering A11 Decision Plot}
    \end{minipage}%
    ~~~
    \begin{minipage}{0.47\textwidth}
        \includegraphics[width=\linewidth]{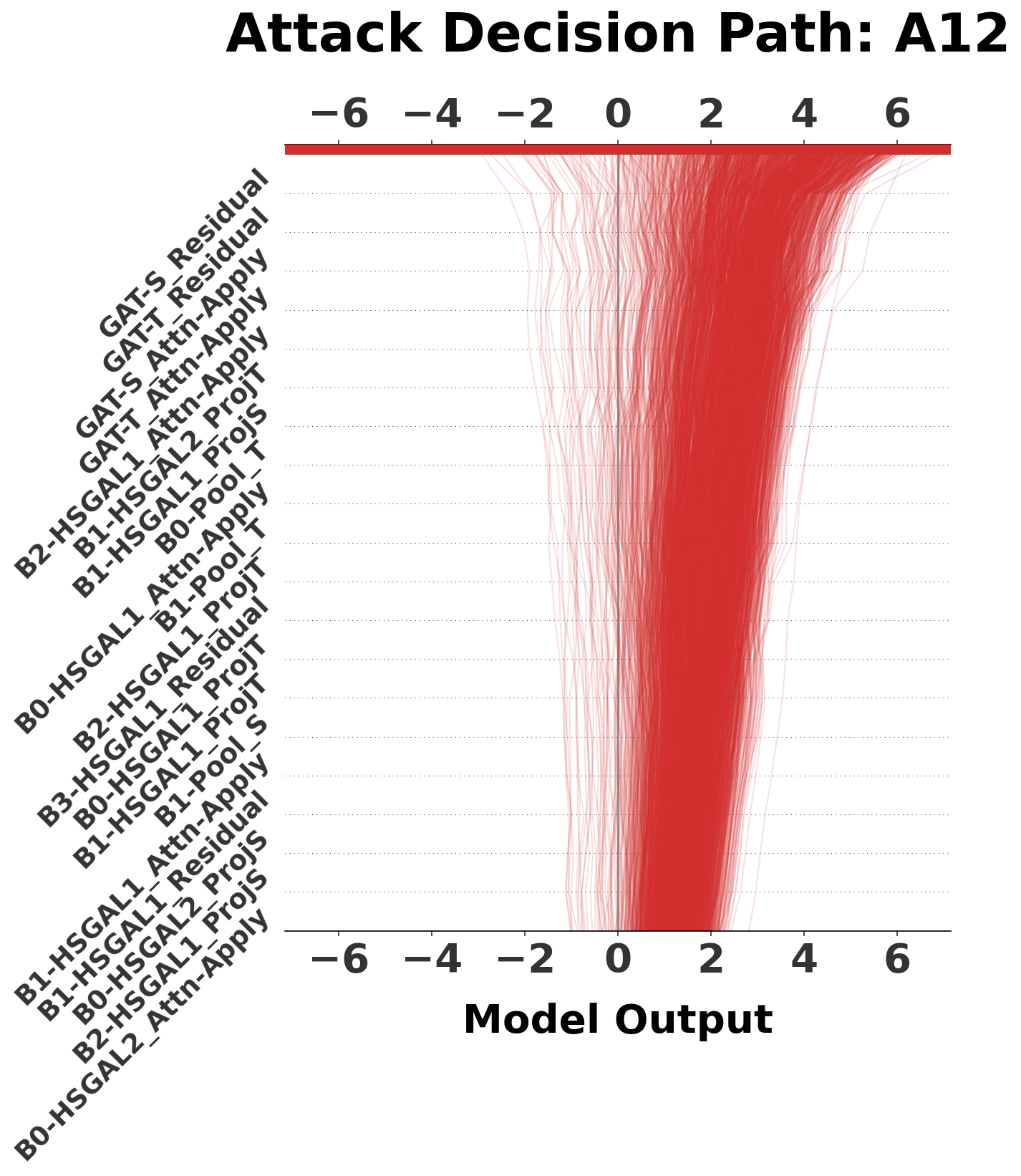}
        \caption*{\centering A12 Decision Plot}
    \end{minipage}
    \caption{Comparative Decision Plots for Attacks A11 (\textbf{Left}) and A12 (\textbf{Right}).}
    \label{fig:dec_a11_a12}
\end{figure}

{For Attack A11, the~model employs Effective Specialization driven by GAT-S ($21.9\%$) and Branch B0 ($20.4\%$), achieving a Confidence Score of $0.76$. Notably, Figure~\ref{fig:shap_a11} demonstrates that the model successfully overcomes a negative (suppressive) contribution from GAT-T to reach a correct decision, highlighting the system's ability to filter out misleading temporal cues in favor of spectral evidence. Attack A12, however, falls into Ineffective Consensus. The~SHAP values (Figure~\ref{fig:shap_a12}) are uniformly low, and~the decision plot (Figure~\ref{fig:dec_a11_a12}) is dominated by noise. This lack of signal is reflected in the metrics (Table~\ref{tab:main_analysis}), where no branch exceeds a 20\% share, indicating that the artifacts generated by the neural waveform model in A12 do not strongly activate any specific architectural component.}

\vspace{-3pt}

\vspace{-6pt}

\vspace{-6pt}

\vspace{-6pt}

\subsubsection{Attacks A13 and A14 (Borderline vs. Distinct)}

\begin{figure}[H]
    \includegraphics[width=.99\linewidth]{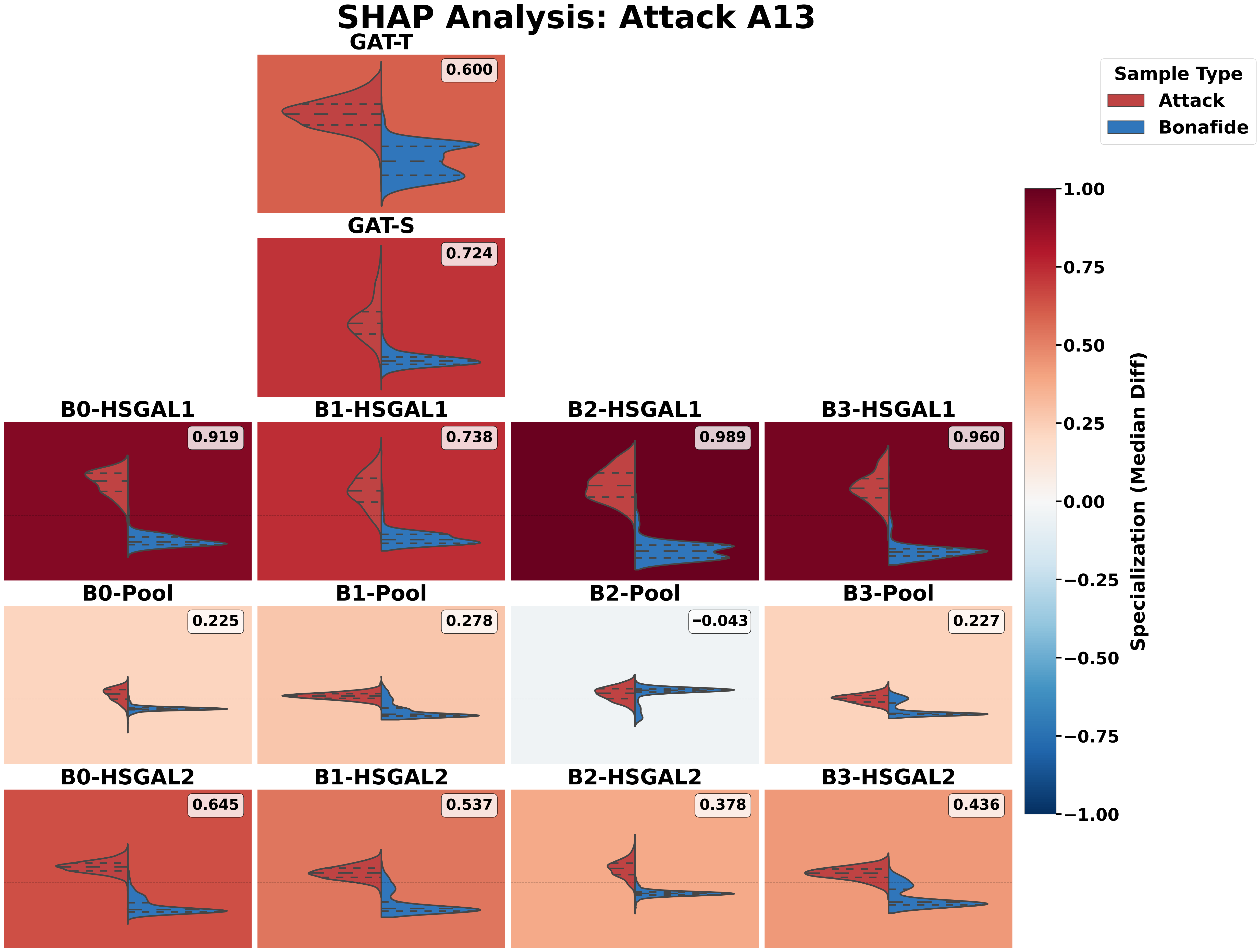}
    \caption{SHAP Distribution for Attack A13. Red and blue distributions represent SHAP values for attack (positive, spoof-indicating) and bonafide (negative, genuine-indicating) samples, respectively.}
    \label{fig:shap_a13}
\end{figure}

\begin{figure}[H]
    \includegraphics[width=.99\linewidth]{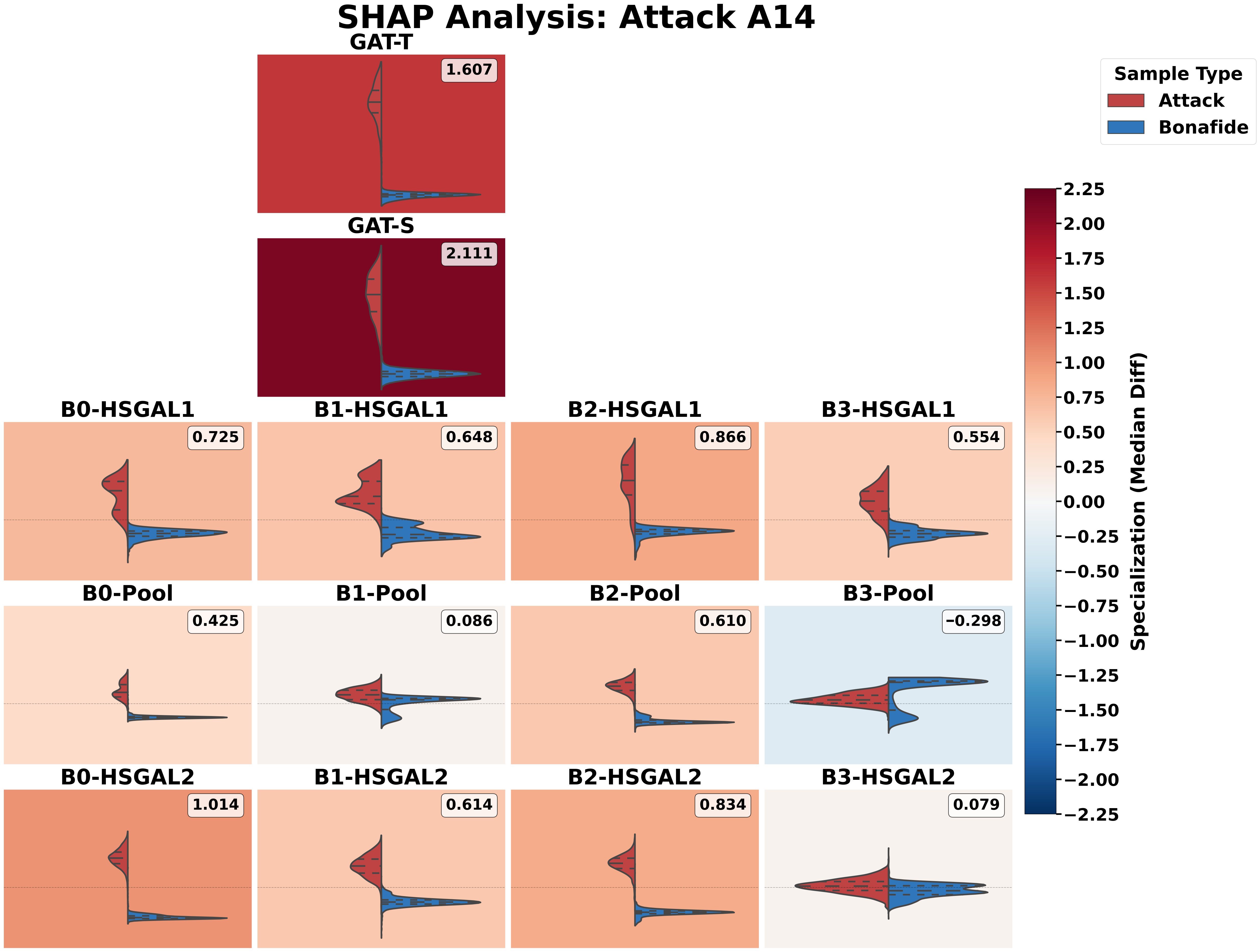}
    \caption{{SHAP} 
 Distribution for Attack A14. Red and blue distributions represent SHAP values for attack (positive, spoof-indicating) and bonafide (negative, genuine-indicating) samples, respectively.}
    \label{fig:shap_a14}
\end{figure}

\begin{figure}[H]
    \begin{minipage}{0.47\textwidth}
        \includegraphics[width=\linewidth]{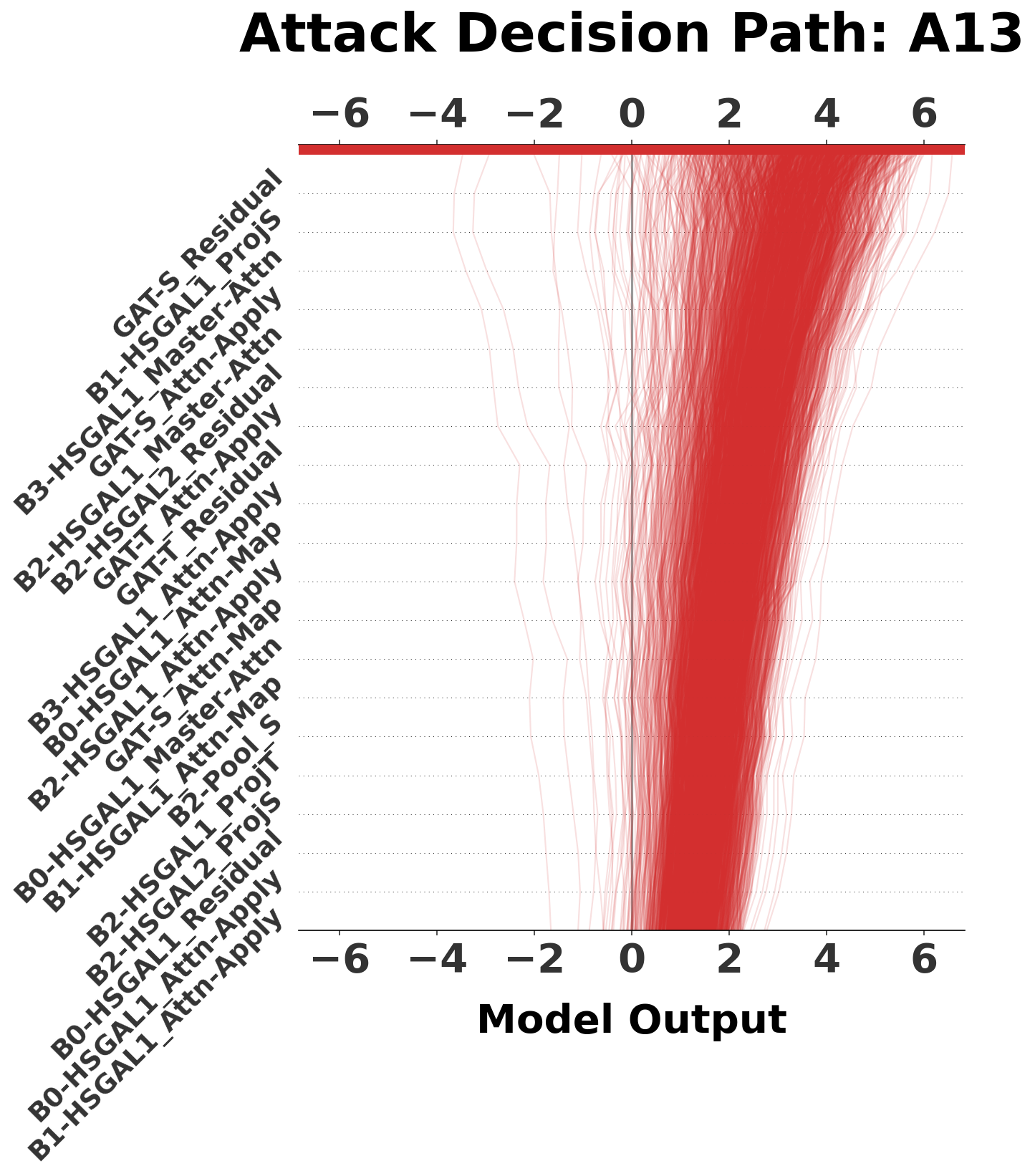}
        \caption*{\centering A13 Decision Plot}
    \end{minipage}%
    ~~~
    \begin{minipage}{0.51\textwidth}
        \includegraphics[width=\linewidth]{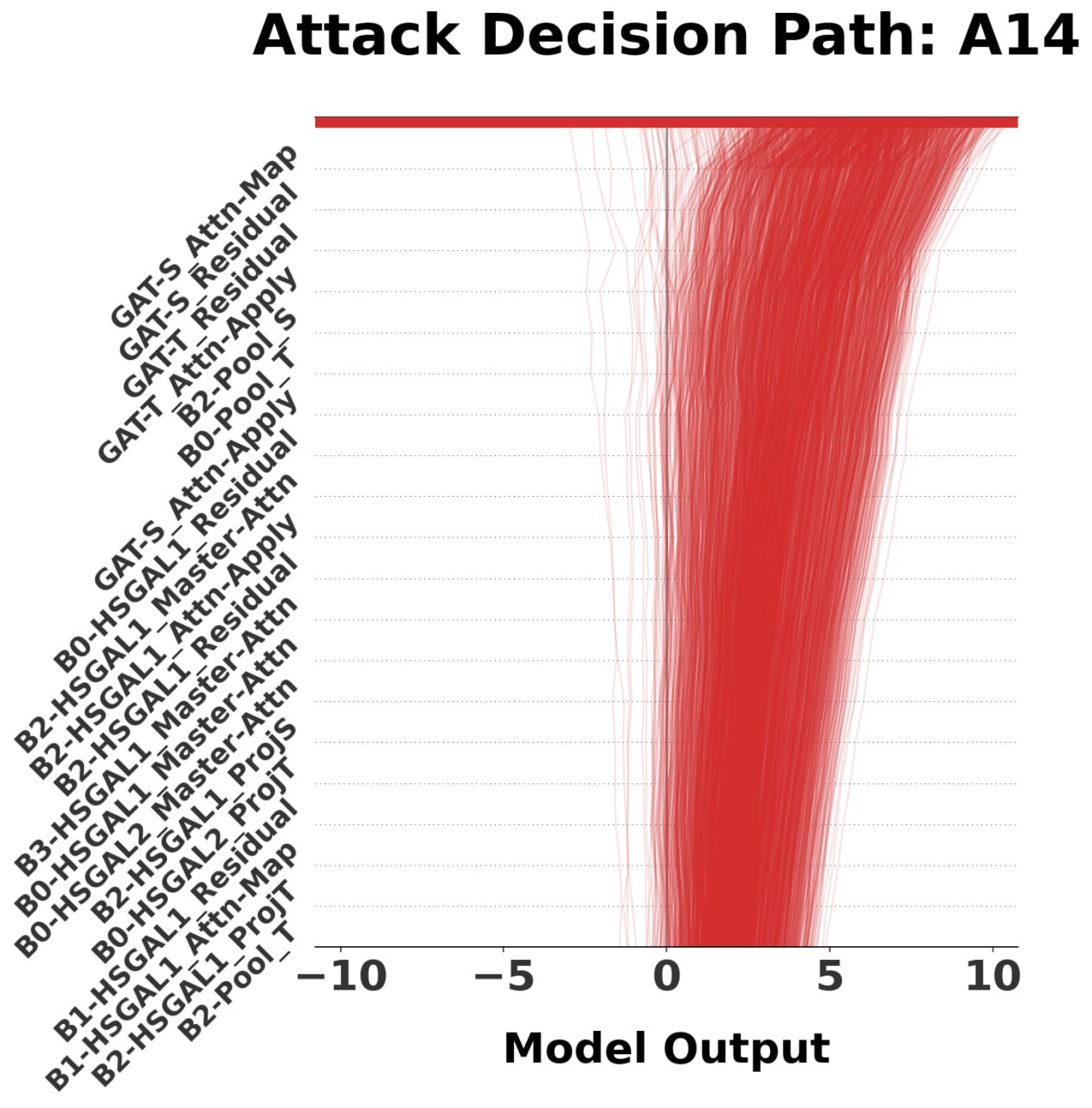}
        \caption*{\centering A14 Decision Plot}
    \end{minipage}
    \caption{Comparative Decision Plots for Attacks A13 (\textbf{Left}) and {A14} 
 (\textbf{Right}).}
    \label{fig:dec_a13_a14}
\end{figure}

{Attack A13 represents a borderline case of Ineffective Consensus (Figure~\ref{fig:shap_a13}); the model aggregates many small signals to reach a decision, yielding a lower Confidence Score of $0.63$. Attack A14 is efficiently detected via Effective Specialization, with~GAT-S ($22.1\%$) and Branch B2 ($20.6\%$) acting as a powerful expert duo ({Figure}
~\ref{fig:shap_a14}). This strong signal results in a high Confidence Score of $1.13$, and~the sharp, unidirectional rise in the decision plot (Figure~\ref{fig:dec_a13_a14}) confirms their dominance. 

This suggests that while the attack is detected, it lies near the decision boundary of the learned feature space, requiring broad architectural participation rather than \mbox{expert delegation.}}

\vspace{-6pt}

\vspace{-6pt}
\vspace{-6pt}

\subsubsection{Attacks A15 and A16 (Weak vs. Robust Consensus)}

\begin{figure}[H]
    \includegraphics[width=.99\linewidth]{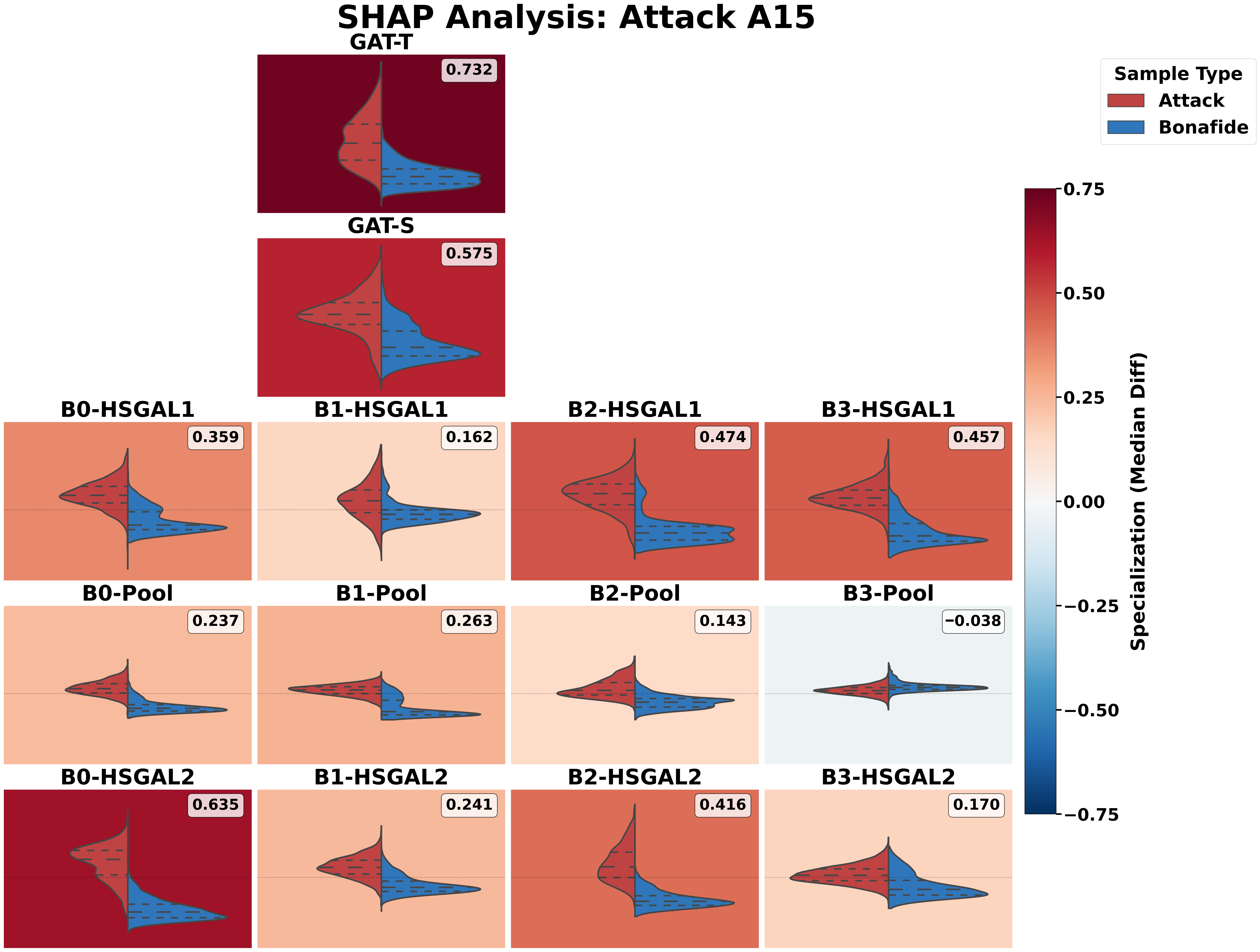}
    \caption{{SHAP} 
 Distribution for Attack A15. Red and blue distributions represent SHAP values for attack (positive, spoof-indicating) and bonafide (negative, genuine-indicating) samples, respectively.}
    \label{fig:shap_a15}
\end{figure}

\begin{figure}[H]
    \includegraphics[width=.99\linewidth]{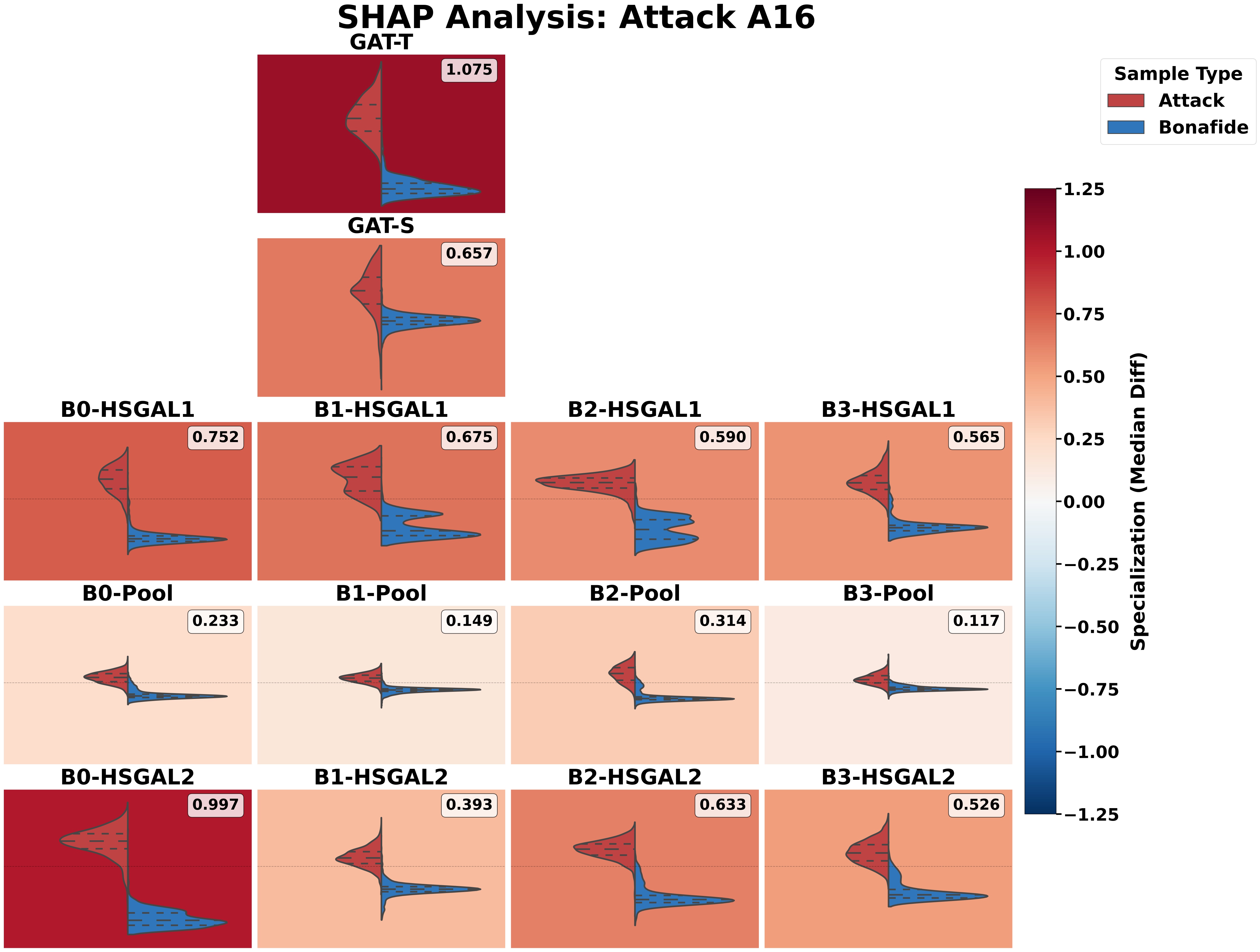}
    \caption{SHAP Distribution for Attack A16. Red and blue distributions represent SHAP values for attack (positive, spoof-indicating) and bonafide (negative, genuine-indicating) samples, respectively.}
    \label{fig:shap_a16}
\end{figure}

\begin{figure}[H]
    \begin{minipage}{0.48\textwidth}
        \includegraphics[width=\linewidth]{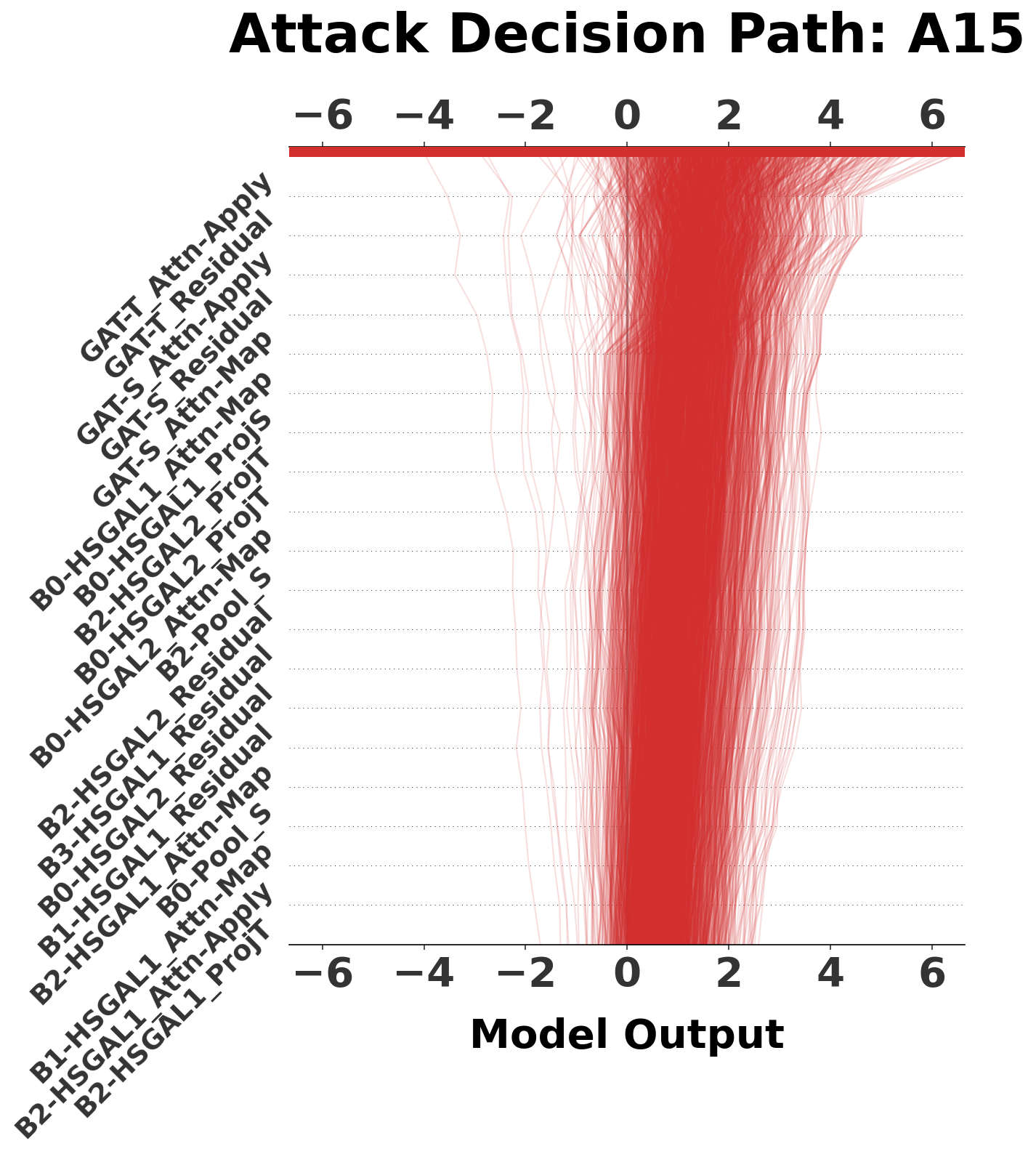}
        \caption*{\centering A15 Decision Plot}
    \end{minipage}%
   ~~~
    \begin{minipage}{0.47\textwidth}
        \includegraphics[width=\linewidth]{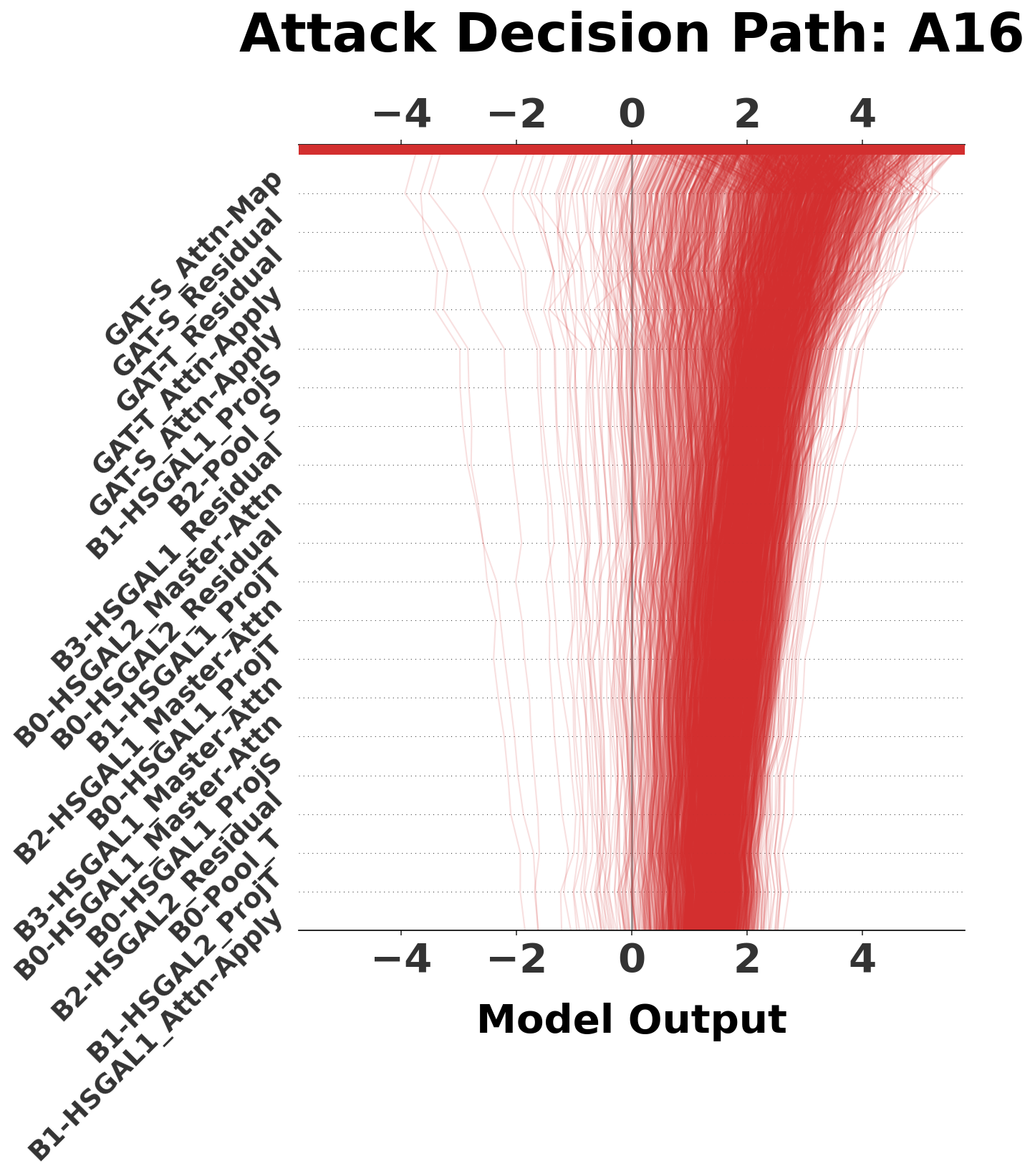}
        \caption*{\centering A16 Decision Plot}
    \end{minipage}
    \caption{{Comparative} 
 Decision Plots for Attacks A15 (\textbf{Left}) and A16 (\textbf{Right}).}
    \label{fig:dec_a15_a16}
\end{figure}

{Attack A15 reveals Ineffective Consensus similar to A08, with~diffuse SHAP distributions (Figure~\ref{fig:shap_a15}) and a very low Confidence Score of $0.38$ (Table~\ref{tab:main_analysis}). Attack A16, conversely, is a robust example of Effective Consensus. Despite being a low-EER attack ($0.72\%$), the~model does not rely on a single branch; instead, Figure~\ref{fig:shap_a16} shows consistent positive contributions across the entire architecture. This distributed strategy maintains a stable Confidence Score of $0.49$ without high variance, providing a highly stable and fault-tolerant detection mechanism (Figure~\ref{fig:dec_a15_a16})}.

\vspace{-3pt}

\vspace{-6pt}

\subsubsection{Attacks A17 and A18 (Vulnerability and Failure)}

\begin{figure}[H]
    \includegraphics[width=.99\linewidth]{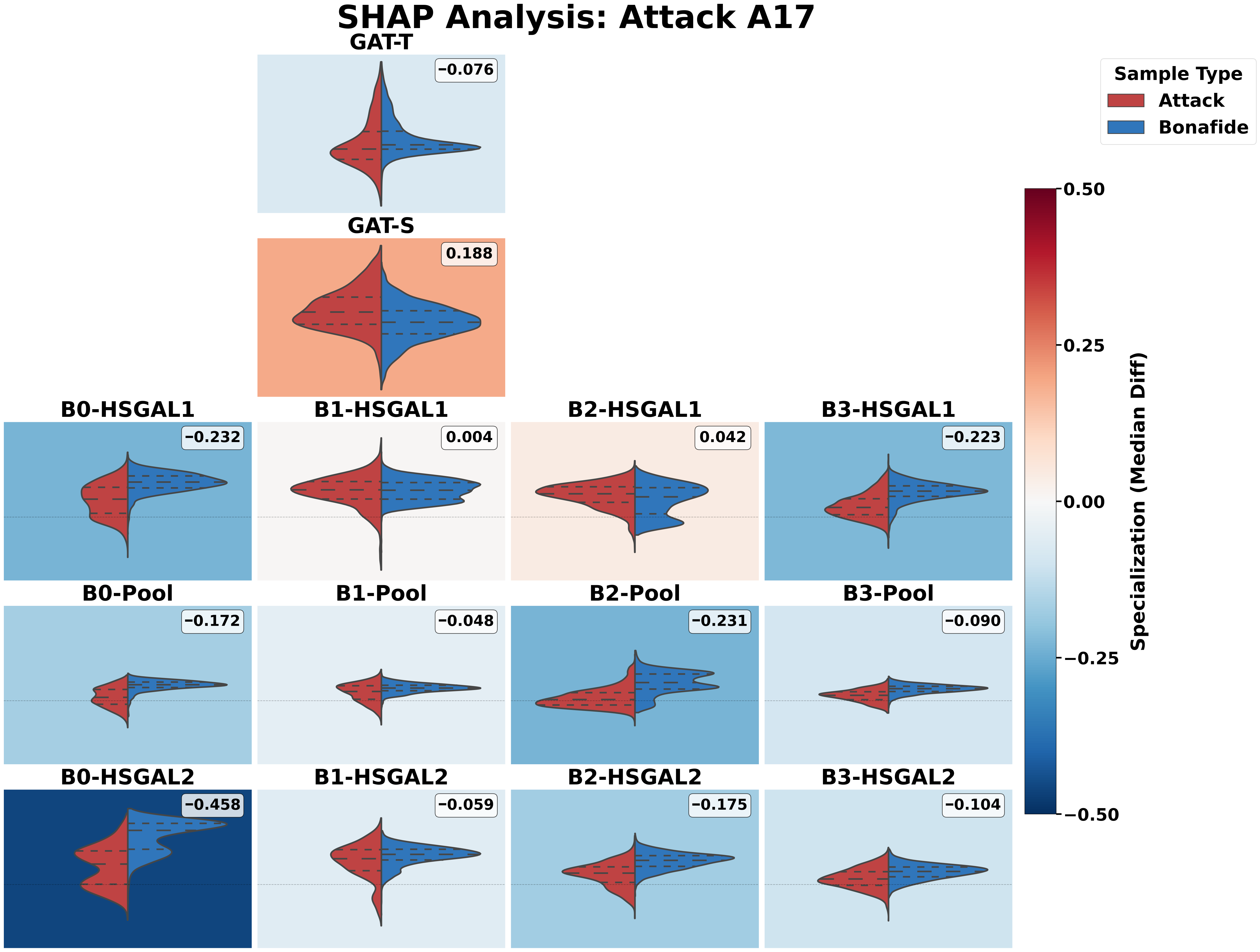}
    \caption{{SHAP} 
 Distribution for Attack A17. Red and blue distributions represent SHAP values for attack (positive, spoof-indicating) and bonafide (negative, genuine-indicating) samples, respectively.}
    \label{fig:shap_a17}
\end{figure}

\begin{figure}[H]
    \begin{minipage}{0.47\textwidth}
        \includegraphics[width=\linewidth]{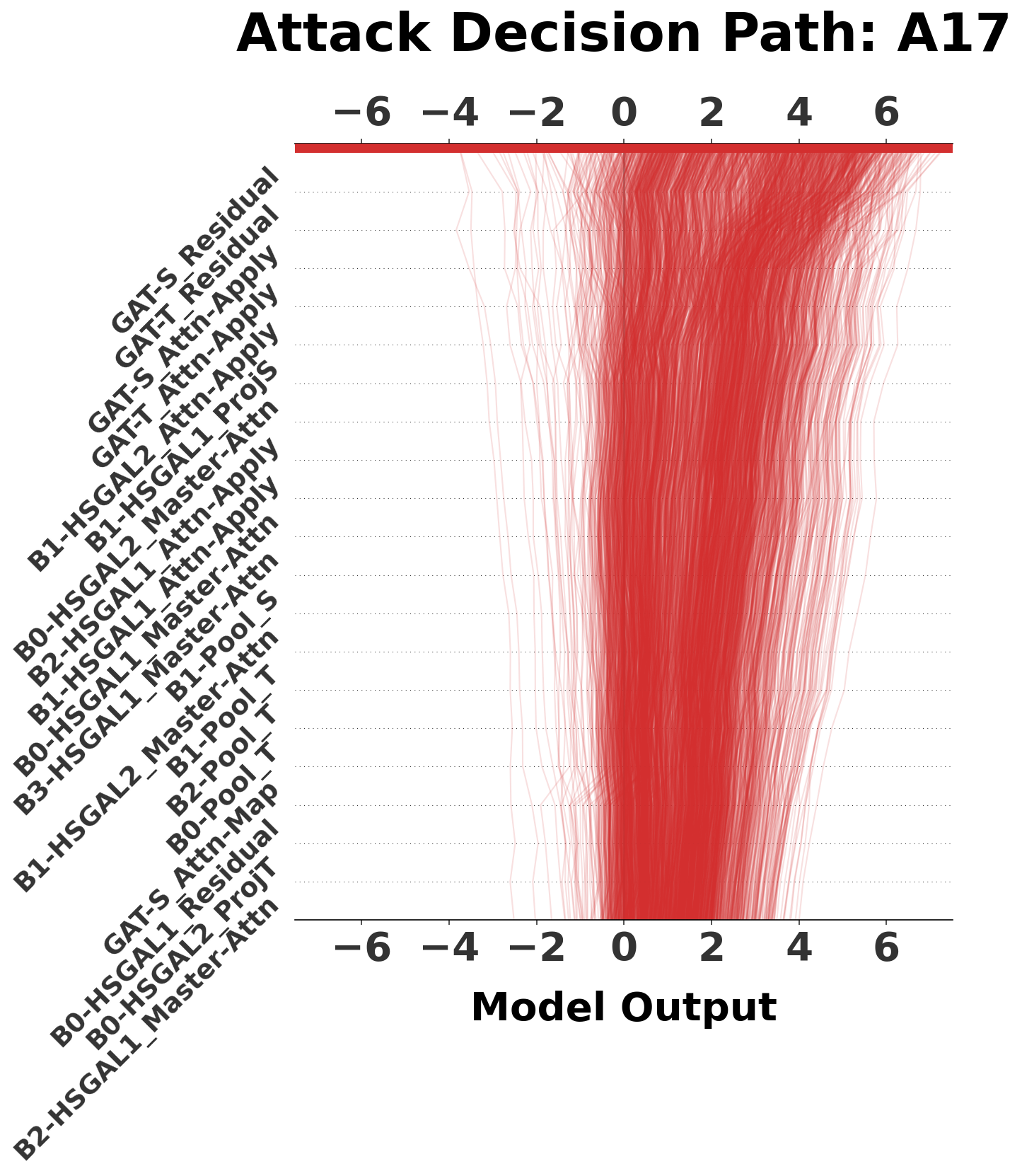}
        \caption*{\centering A17 Decision Plot}
    \end{minipage}%
    ~~~
    \begin{minipage}{0.48\textwidth}
        \includegraphics[width=\linewidth]{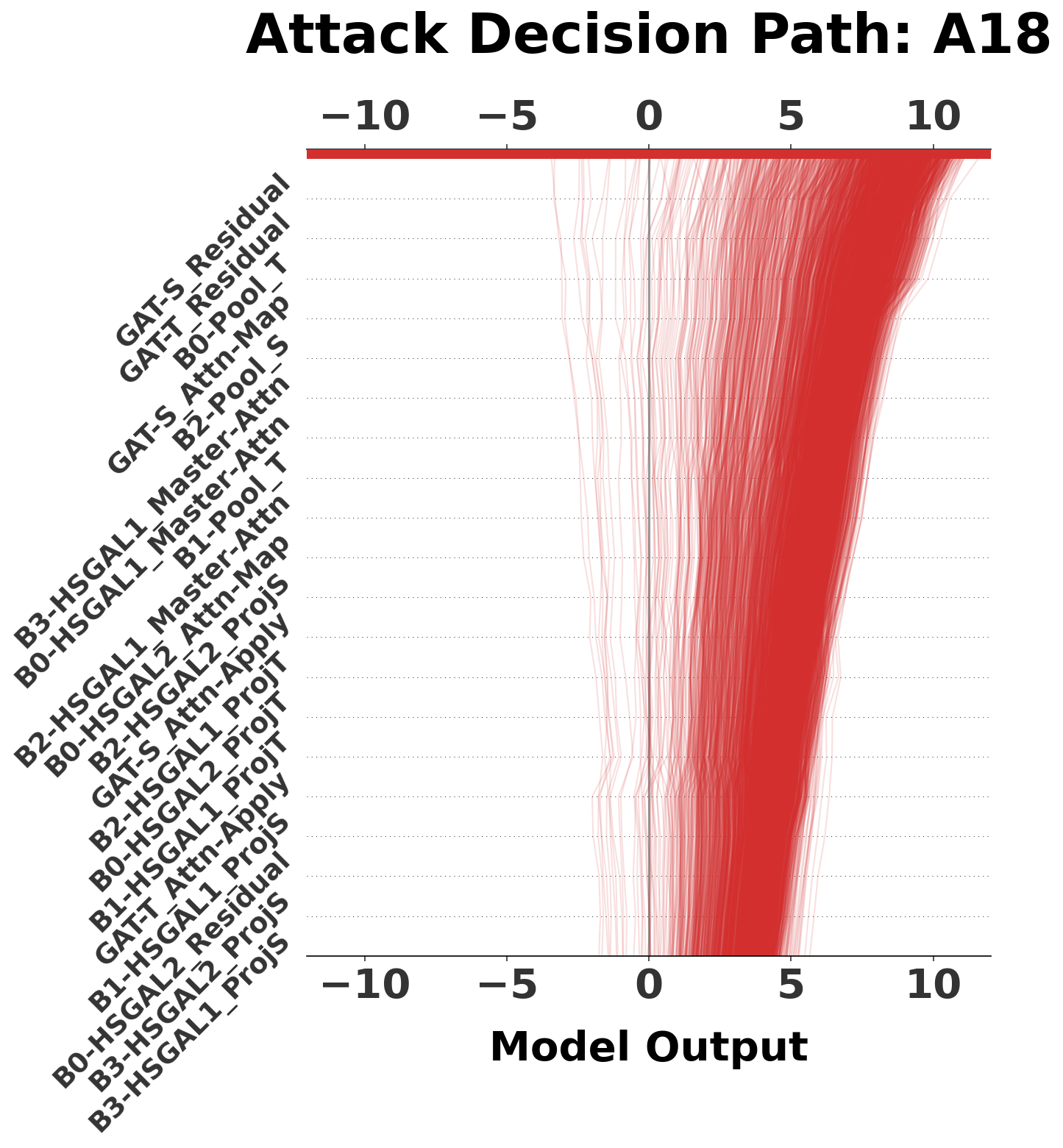}
        \caption*{\centering A18 Decision Plot}
    \end{minipage}
    \caption{{Comparative} 
 Decision Plots for Attacks A17 (\textbf{Left}) and A18 (\textbf{Right}).}
    \label{fig:dec_a17_a18}
\end{figure}

\begin{figure}[H]
    \includegraphics[width=.99\linewidth]{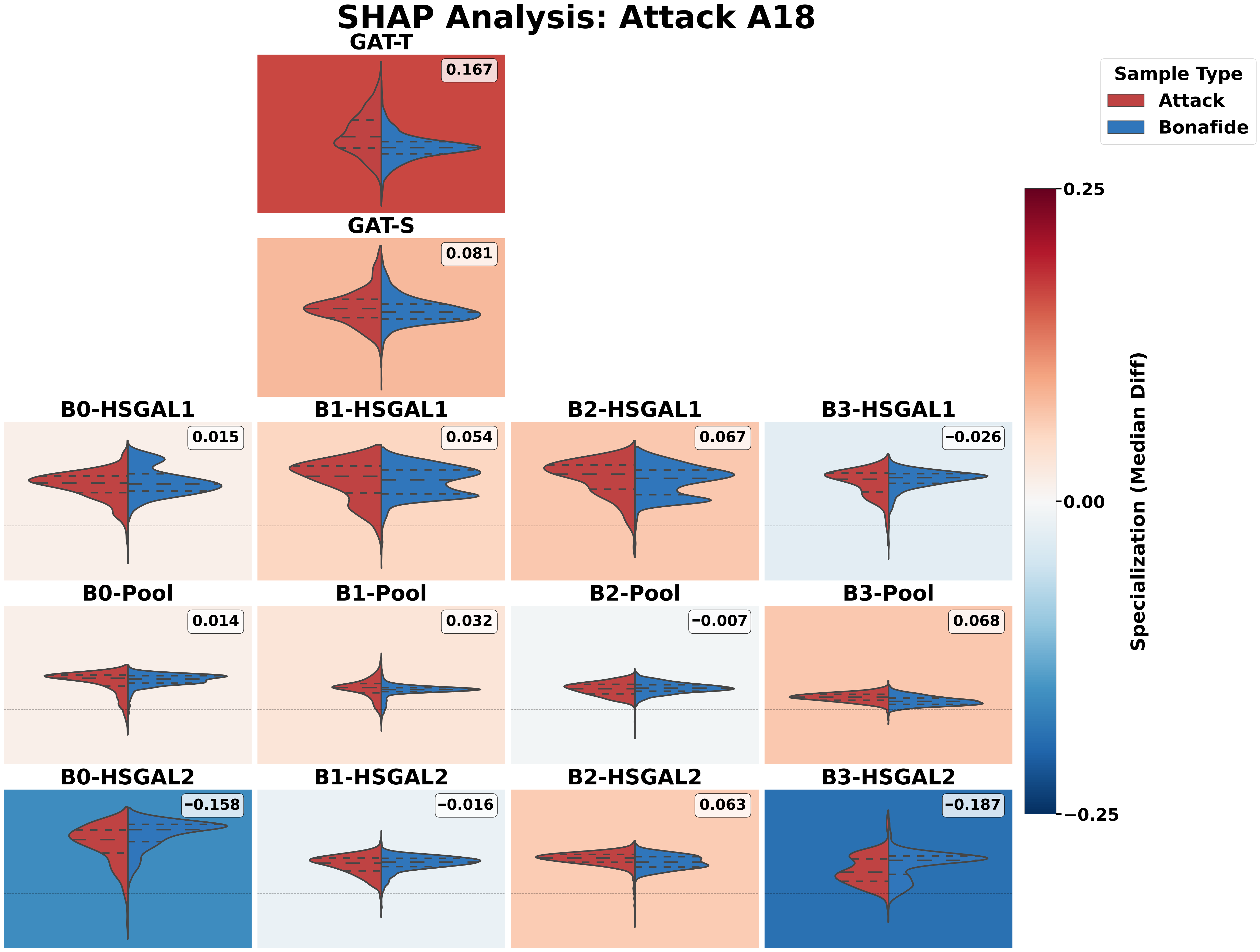}
    \caption{{SHAP} 
 Distribution for Attack A18. Red and blue distributions represent SHAP values for attack (positive, spoof-indicating) and bonafide (negative, genuine-indicating) samples, respectively.}
    \label{fig:shap_a18}
\end{figure}

{These attacks expose the Flawed Specialization vulnerability. For~Attack A17 (Figure~\ref{fig:shap_a17}, the~model confidently relies on Branch B1 ($20.9\%$ share), yet the high EER ($14.26\%$) indicates this confidence is misplaced. The~failure is most  high-error for Attack A18, where the model delegates the decision to Branch B0 (25.0\% share, Table~\ref{tab:main_analysis}). The~decision plot ({Figure}
~\ref{fig:dec_a17_a18}) illustrates this pathology: a strong, confident step driven by the wrong features leads to a systematic misclassification (Figure~\ref{fig:shap_a18}). This behavior is distinct from the low-confidence confusion of A08; here, the~Confidence Score is relatively high ($1.33$ for A18), quantitatively proving that the model is confidently wrong rather than uncertain.}

\vspace{-6pt}

\vspace{-6pt}

\vspace{-6pt}

\subsubsection{Attack A19 (Global Spectral Detection)}

\begin{figure}[H]
    \includegraphics[width=0.45\linewidth]{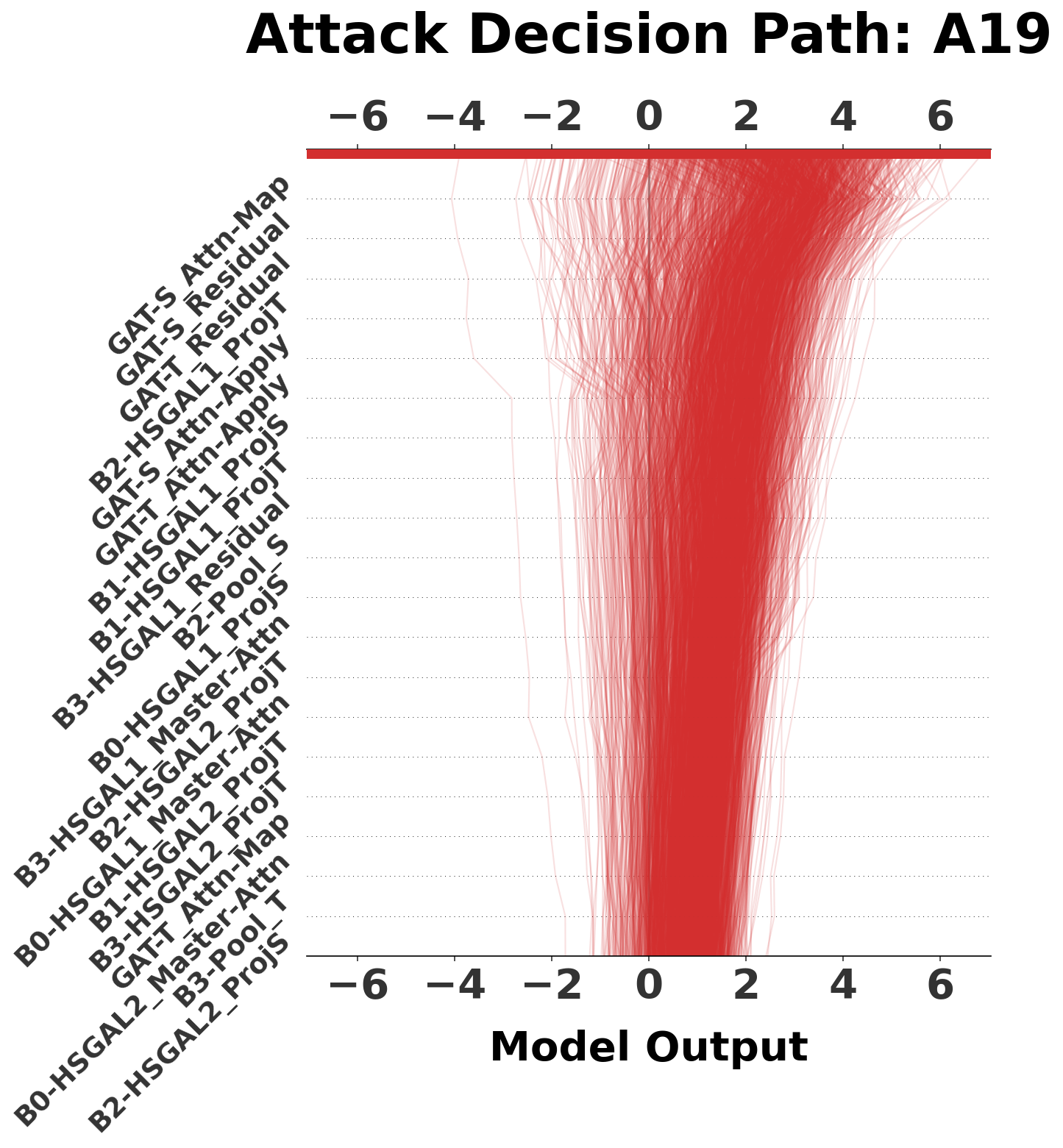}
    \caption{{Decision} 
 Plot for Attack~A19.}
    \label{fig:dec_a19}
\end{figure}

\begin{figure}[H]
    \includegraphics[width=.99\linewidth]{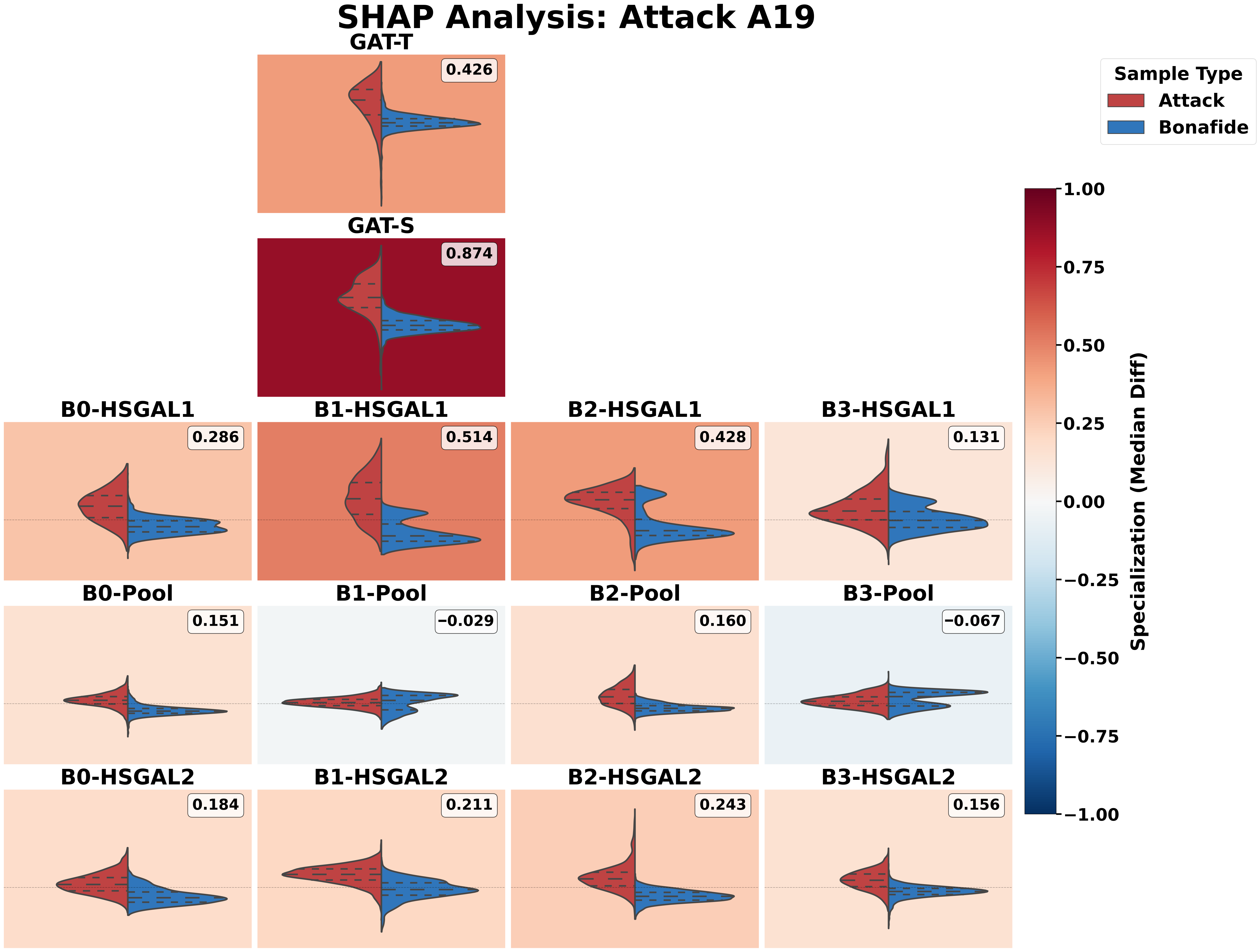}
    \caption{SHAP Distribution for Attack A19. Red and blue distributions represent SHAP values for attack (positive, spoof-indicating) and bonafide (negative, genuine-indicating) samples, respectively.}
    \label{fig:a19_shap}
\end{figure}

{Attack A19 is resolved via Effective Specialization led by the global GAT-S module (19.5\% share). The~decision trajectories presented on Figure~\ref{fig:dec_a19} confirm that the final decision pivots on the contribution of the global module. Although~the Confidence Score is low ($0.45$), {Figure}
~\ref{fig:a19_shap} shows that the global spectral features provide the decisive signal, effectively overriding minor disagreements from the local pooling layers. This validates the architectural choice of including global attention modules alongside local processing branches. }


\vspace{-6pt}

\subsection{Detailed Statistical~Data}
For completeness, this subsection provides the detailed numerical data used for the analysis. Table~\ref{tab:appendix_shap_sum} shows the summed mean SHAP values for each of the six main architectural components. Table~\ref{tab:appendix_softmax} shows the final calculated contribution~shares.

\vspace{-6pt}



\subsection{Single-Branch Retention~Ablation}

{To complement the descriptive SHAP-based strategy analysis, a~small functional ablation was conducted to test whether the dominant branch identified by the attribution pipeline is sufficient to preserve detection performance on difficult attacks. We focus on the three highest-error cases in our evaluation (A10, A17, A18), where the model either exhibits ineffective consensus/specialization (A10) or flawed specialization (A17, A18) in Table~\ref{tab:main_analysis}. For~each selected attack, we retain only the most dominant branch and set the activations of all other branches/modules to zero during inference, after~which EER is recomputed on the same evaluation protocol.}

{The results in Table~\ref{tab:retention_ablation} show that restricting AASIST3 to a single “best” branch strongly degrades performance, increasing EER to 63–68\% for all tested attacks.This behavior is consistent with the interpretation that our SHAP-based dominance scores quantify relative reliance patterns under the full architecture, rather than implying that the dominant branch alone implements a complete and causally sufficient detector.
In particular, for~A17 and A18 (flawed specialization), the~model’s high reliance on B1/B0 already corresponds to a failure mode under the full system (Table~\ref{tab:main_analysis}), and~isolating the same branch removes any possibility of corrective evidence from alternative branches, further amplifying the error.
}

\begin{table}[H]
\centering
\caption{Single-Branch Ablation for High-Error Attacks. EER when retaining only the dominant branch identified in Table~\ref{tab:main_analysis}, with~all other branches~zeroed.}
\label{tab:retention_ablation}
\begin{tabularx}{\textwidth}{llcc}
\toprule
\textbf{Attack} & \textbf{Dominant Branch Kept} & \textbf{Baseline EER} & \textbf{EER with Only Dominant Branch} \\ \midrule

A18 &	B0 & 	28.63 & 63.34\\
A17	&B1&	14.26 &66.81 \\
A10	&B2&	17.31 &	67.55\\
\bottomrule
\end{tabularx}
\end{table}
\unskip

\section{Discussion}
\label{sec:discussion}

The results presented in Section~\ref{sec:results} allow for a deep interpretation of the AASIST3~\cite{Borodin2024aasist3} model's internal operating logic. By~correlating the model's chosen operational strategy with its empirical performance (EER), we can move beyond simple descriptions of feature importance to a diagnostic analysis of architectural strengths and~weaknesses.

\subsection{Analysis of Operational~Archetypes}
The correlation analysis confirms that the model dynamically switches between the four operational archetypes defined in our methodology. We categorize the behavior of the model on all 13 analyzed attacks (A07--A19) as follows:

 {Effective Specialization:} Our findings demonstrate that the architecture is highly efficient when it can isolate distinct artifacts. This group includes attacks  {A09},  {A14},  {A11}, and~ {A19}. By~successfully delegating detection to a single ``expert'' component (e.g., Branch B2 for A09 or GAT-S for A11), the~model filters out noise from less relevant branches. This suggests that for these ``easy'' attacks, the~multi-branch structure functions primarily as a feature selection~mechanism.

 {Effective Consensus:} The low EERs observed for attacks  {A07} and  {A16} validate the redundancy hypothesis of multi-branch networks. In~these cases, the~artifacts appear to be broad enough to be captured by disparate feature extractors (temporal and spectral). The~agreement across branches acts as a robust error-correction mechanism, ensuring stability even if one branch provides a noisy~signal.

 {Ineffective Consensus:} This behavior reveals a representational limit of the current architecture and is observed for attacks  {A08},  {A12},  {A13}, and~ {A15}. The~balanced but weak contributions indicate that the model is ``confused'' rather than ``collaborative.'' The lack of a strong signal from any branch suggests that the specific artifacts of these attacks fall into the blind spots of the learned feature filters in all four branches~simultaneously.

 {Flawed Specialization:} This archetype represents the most important discovery of our study. Attacks  {A18} and  {A17} fall into this category, as~does  {A10} (which exhibits ineffective specialization). For~A18 and A17, high internal confidence does not equate to accuracy. The~model's tendency to ``lock on'' to a misleading feature in a single branch (e.g., B0 for A18) without cross-verification from other branches points to a lack of inter-branch regularization. Attack A10 similarly shows reliance on specific components (GAT-T) but fails to achieve accuracy, highlighting that specialization is detrimental when the selected expert is~incompetent.

\subsection{Architectural Implications and~Vulnerabilities}
The identification of these behaviors has direct implications for understanding the model's robustness and guiding future~improvements.

First, as~detailed in Table~\ref{tab:appendix_softmax}, while effective specialization is efficient, it introduces  {single points of failure}. The~model's heavy reliance on specific branches—for example, Branch B2 contributes 25.5\% of the decision for A09, and~GAT-S contributes 22.1\% for A14—implies that an adversarial attack specifically optimizing against these branches would likely bypass the entire system. Other branches, having low contribution shares (often below 10--15\%), would be unable to~compensate.

Second, the~ {Flawed Specialization} pattern highlights a missing meta-cognitive component in the architecture. The~current design aggregates features but does not seemingly evaluate the {reliability} of a branch's high activation against the consensus of others. A~robust system should be able to penalize a high-confidence branch (like B1 in A17 with 20.9\% share) if it contradicts the collective disagreement of the other three~branches.

Finally, the~roles of the global modules are distinct. The~GAT-S (spectral) module acts as a consistent, high-performance anomaly detector, frequently leading effective strategies (e.g., A11, A19). In~contrast, the~GAT-T (temporal) module appears less reliable as a standalone detector, suggesting that temporal artifacts in the ASVspoof dataset are either more subtle or harder to disentangle from bonafide speech~dynamics.

{The observed inferiority of GAT-T relative to GAT-S, particularly its failure to specialize under sophisticated neural synthesis attacks like A10 (Tacotron 2 with WaveRNN), can be attributed to the acoustic properties of the ASVspoof 2019 logical access dataset~\citep{wang2020asvspoof2019largescalepublic}. Neural vocoders typically operate by predicting audio samples or spectral frames sequentially, a~process that inherently preserves long-range temporal continuity and prosodic consistency, thereby producing temporal envelopes that closely mimic bona fide speech~\citep{8461368, Siuzdak2024Vocos}. In~contrast, the~generation process frequently introduces distinct spectral artifacts—such as overly smoothed formants, phase discontinuities, or~checkerboard patterns in the mel-spectrogram—arising from the frame-wise reconstruction loss or upsampling operations~\citep{9414913}. Consequently, the~discriminatory signal is far denser in the frequency domain, rendering the spectral graph attention (GAT-S) a more robust expert for artifact detection, whereas the temporal graph attention (GAT-T) struggles to find reliable distinguishing features in the comparatively well-modeled time domain~\cite{tak21_asvspoof}.
}

\subsection{Connection to Broader~Research}
Our findings empirically support theories from the wider field of deep learning. The~dynamic adoption of specialized roles for specific tasks confirms observations that multi-branch networks naturally diversify to cover different feature subspaces~\cite{Zhang2024multislice}. However, our identification of ``Flawed Specialization'' adds nuance to the work on uncertainty estimation~\cite{benartzy2024attend}: we show that model failure is not always preceded by high uncertainty (entropy); often, the~most  high-error failures occur when the model is irrationally certain about a wrong~feature.

{Although the empirical analysis in this work is restricted to the AASIST3 architecture~\cite{Borodin2024aasist3}, the~proposed spectral–SHAP framework is, by~construction, model-agnostic and can in principle be applied to other mainstream multi-branch anti-spoofing systems, including RawNet-style encoders~\citep{9414234}, graph-based countermeasures~\citep{tak21_asvspoof}, and~recent convolutional–transformer hybrids~\citep{rosello2023conformer} used in ASVspoof challenges~\cite{wang24_asvspoof}. The~only requirements are access to intermediate activations of a finite set of branches or experts and the ability to compute covariance operators over these activations, so that spectral signatures and SHAP-based contribution shares can be defined in the same manner as for AASIST3. This makes the framework suitable not only for classical parallel-branch countermeasures but also for modern mixture-of-experts~\citep{dalterio2025attention} and gating-based anti-spoofing models~\citep{tran2025multilevel}, where each expert or specialist sub-network can be treated as a branch and analyzed in terms of its operational strategy, redundancy, and~potential single points of failure.}

\section{Conclusions}
\label{sec:conclusion}

This study presented a rigorous methodology for interpreting the internal decision-making strategies of the multi-branch AASIST3 architecture. By~correlating spectral feature attributions with the Equal Error Rate (EER) across 13 distinct spoofing attacks, we moved beyond black-box performance metrics to a granular understanding of \mbox{architectural behavior}.

Our analysis revealed that AASIST3 is not a static feature extractor but a dynamic system that adopts different operational strategies based on the input. We identified that for ``easy'' attacks (e.g., A09, A14), the~model effectively utilizes a  {Specialization} strategy, delegating detection to a specific branch (often B2 or GAT-S) that filters out noise. Conversely, for~broadly detectable attacks (e.g., A07, A16), it employs an  {Effective Consensus} strategy, leveraging redundancy to ensure~stability.

Crucially, our framework exposed a significant architectural vulnerability:  {Flawed Specialization}. In~the most difficult attack scenarios (A17, A18), the~model exhibited high internal confidence in a single branch (e.g., B0 for A18), while other branches provided weak or conflicting signals. This misplaced confidence led to  high-error performance (EER > 28\%), identifying a notable lack of inter-branch regularization and cross-verification mechanisms. Furthermore, the~identification of ``Single Points of Failure'' in successful specialization scenarios suggests that while efficient, the~current architecture may be susceptible to adversarial attacks targeting specific ``expert''~branches.

\vspace{6pt} 

\authorcontributions{Conceptualization, I.V. and K.B.; methodology, I.V. and K.B.; software, I.V.; validation, I.V., K.B. and G.M.; formal analysis, I.V. and K.B.; investigation, I.V. and K.B.; resources, I.V.; data curation, I.V.; writing—original draft preparation, I.V. and K.B.; writing—review and editing, I.V., K.B., M.G. and G.M.; visualization, I.V.; supervision, K.B. and G.M.; project administration, K.B., M.G. and G.M.; funding acquisition, G.M. All authors have read and agreed to the published version of the manuscript.}

\funding{This research received no external~funding.}

\institutionalreview{{Not applicable.} 
}
\informedconsent{{Not applicable.} 
}
\dataavailability{
\textls[-10]{{The original data} 
 presented in the study are openly available in DataShare, HuggingFace GitHub at \url{https://datashare.ed.ac.uk/handle/10283/3336} (accessed on 10 November 2025)  {(ASVSpoof2019)}
, \url{https://huggingface.co/lab260/AASIST3} accessed on 10 November 2025) {(AASIST3)} and}  \url{https://github.com/lab260ru/Interpreting-Multi-Branch-Anti-Spoofing-Architectures} {(our code)}{.} 

}

\conflictsofinterest{The authors declare no conflicts of~interest.} 

\abbreviations{Abbreviations}{
The following abbreviations are used in this manuscript:
\\

\noindent 
\begin{tabular}{@{}ll}
AASIST & Audio Anti-Spoofing using Integrated Spectro-Temporal GNNs \\
EER & Equal Error Rate \\
GAT & Graph Attention Network \\
HSGAL & Heterogeneous Stacking Graph Attention Layers \\
SHAP & SHapley Additive exPlanations \\
\end{tabular}
}

\begin{adjustwidth}{-\extralength}{0cm}
\reftitle{References}

\begin{thebibliography}{999}

\bibitem[Yamagishi et~al.(2021)Yamagishi, Wang, Todisco, Sahidullah, Patino,
  Nautsch, Liu, Lee, Kinnunen, Evans, and Delgado]{asvspoof2021}
{Yamagishi, J.;} 
 Wang, X.; Todisco, M.; Sahidullah, M.; Patino, J.; Nautsch, A.;
  Liu, X.; Lee, K.A.; Kinnunen, T.; Evans, N.;  et~al.
\newblock ASVspoof 2021: Accelerating progress in spoofed and deepfake speech
  detection.
\newblock In Proceedings of the 2021 Edition of the Automatic Speaker
  Verification and Spoofing Countermeasures Challenge, {Online, 16 September} 
2021; pp. 47--54.
\newblock {\url{https://doi.org/10.21437/ASVSPOOF.2021-8}}.

\bibitem[Wang et~al.(2024)Wang, Delgado, Tak, weon Jung, jin Shim, Todisco,
  Kukanov, Liu, Sahidullah, Kinnunen, Evans, Lee, and Yamagishi]{asvspoof5}
{Wang, X.; Delgado, H.; Tak, H.; Jung, J.-W.; Shim, H.-J.; Todisco, M.;} 
  Kukanov, I.; Liu, X.; Sahidullah, M.; Kinnunen, T.H.;  et~al.
\newblock ASVspoof 5: Crowdsourced speech data, deepfakes, and adversarial
  attacks at scale.
\newblock In Proceedings of the Automatic Speaker Verification Spoofing
  Countermeasures Workshop (ASVspoof 2024), {Kos, Greece, 31 August} 2024; pp. 1--8.
\newblock {\url{https://doi.org/10.21437/ASVspoof.2024-1}}.

\bibitem[Wang et~al.(2020)Wang, Yamagishi, Todisco, Delgado, Nautsch, Evans,
  Sahidullah, Vestman, Kinnunen, Lee, Juvela, Alku, Peng, Hwang, Tsao, Wang,
  Maguer, Becker, Henderson, Clark, Zhang, Wang, Jia, Onuma, Mushika, Kaneda,
  Jiang, Liu, Wu, Huang, Toda, Tanaka, Kameoka, Steiner, Matrouf, Bonastre,
  Govender, Ronanki, Zhang, and Ling]{wang2020asvspoof2019largescalepublic}
Wang, X.; Yamagishi, J.; Todisco, M.; Delgado, H.; Nautsch, A.; Evans, N.;
  Sahidullah, M.; Vestman, V.; Kinnunen, T.; Lee, K.A.;  et~al.
\newblock ASVspoof 2019: A large-scale public database of synthesized,
  converted and replayed speech.
\newblock {\em Comput. Speech Lang.} {\bf 2020}, {\em 64},~101114.
\newblock {\url{https://doi.org/https://doi.org/10.1016/j.csl.2020.101114}}.

\bibitem[Borodin et~al.(2024{\natexlab{a}})Borodin, Kudryavtsev, Mkrtchian, and
  Gorodnichev]{Borodin2024ResCapsRes2TCNGuard}
Borodin, K.; Kudryavtsev, V.; Mkrtchian, G.; Gorodnichev, M.
\newblock Capsule-based and TCN-based Approaches for Spoofing Detection in
  Voice Biometry.
\newblock {\em Eng. Technol. Appl. Sci. Res.} {\bf 2024},
  {\em 14},~18409--{18414}
  . 
\newblock {\url{https://doi.org/10.48084/etasr.8906}}.

\bibitem[Borodin et~al.(2024{\natexlab{b}})Borodin, Kudryavtsev, Korzh,
  Efimenko, Mkrtchian, Gorodnichev, and Rogov]{Borodin2024aasist3}
Borodin, K.; Kudryavtsev, V.; Korzh, D.; Efimenko, A.; Mkrtchian, G.;
  Gorodnichev, M.; Rogov, O.Y.
\newblock {AASIST3: KAN-Enhanced AASIST Speech Deepfake Detection using SSL
  Features and Additional Regularization for the ASVspoof 2024 Challenge}. \emph{arXiv} \textbf{2024}, {arXiv:2408.17352.} 

\bibitem[Kinnunen et~al.(2019)Kinnunen, Lee, Delgado, Evans, Todisco,
  Sahidullah, Yamagishi, and Reynolds]{tdcf}
Kinnunen, T.; Lee, K.A.; Delgado, H.; Evans, N.; Todisco, M.; Sahidullah, M.;
  Yamagishi, J.; Reynolds, D.A.
\newblock t-DCF: A Detection Cost Function for the Tandem Assessment of
  Spoofing Countermeasures and Automatic Speaker Verification. \emph{arXiv} \textbf{2019}, arXiv:1804.09618. 

\bibitem[Sundararajan et~al.(2017)Sundararajan, Taly, and
  Yan]{Integrated_gradients}
Sundararajan, M.; Taly, A.; Yan, Q.
\newblock Axiomatic attribution for deep networks.
\newblock In Proceedings of the 34th International
  Conference on Machine Learning, ICML'17, {Sydney, Australia, 6--11 August} 2017; Volume 70, {pp}
. 3319–3328.

\bibitem[Jung et~al.(2022)Jung, Heo, Tak, Shim, Chung, Lee, Yu, and
  Evans]{aasist}
Jung, J.W.; Heo, H.S.; Tak, H.; Shim, H.J.; Chung, J.S.; Lee, B.J.; Yu, H.J.;
  Evans, N.
\newblock AASIST: Audio Anti-Spoofing Using Integrated Spectro-Temporal Graph
  Attention Networks.
\newblock In Proceedings of the ICASSP 2022---2022 IEEE International
  Conference on Acoustics, Speech and Signal Processing (ICASSP), {Singapore, 23--27 May} 2022; pp.
  6367--6371.
\newblock {\url{https://doi.org/10.1109/ICASSP43922.2022.9747766}}.

\bibitem[Hu and Sompolinsky(2022)]{HuSompolinsky2022CovSpectrum}
Hu, Y.; Sompolinsky, H.
\newblock The spectrum of covariance matrices of randomly connected recurrent
  neuronal networks with linear dynamics.
\newblock {\em PLoS Comput. Biol.} {\bf 2022}, {\em 18},~e1010327.
\newblock {\url{https://doi.org/10.1371/journal.pcbi.1010327}}.

\bibitem[Sihag et~al.(2023)Sihag, Mateos, McMillan, and
  Ribeiro]{sihag2023covarianceneuralnetworks}
Sihag, S.; Mateos, G.; McMillan, C.; Ribeiro, A.
\newblock coVariance Neural Networks. \emph{arXiv} \textbf{2023}, arXiv:2205.15856.

\bibitem[Binkowski et~al.(2025)Binkowski, Janiak, Sawczyn, Gabrys, and
  Kajdanowicz]{binkowski2025hallucination}
Binkowski, J.; Janiak, D.; Sawczyn, A.; Gabrys, B.; Kajdanowicz, T.
\newblock Hallucination Detection in LLMs Using Spectral Features of Attention
  Maps. \emph{arXiv} \textbf{2025}, arXiv:2502.17598.
  
  

\bibitem[Harzli and Grau(2025)]{harzli2025adversarial}
Harzli, O.E.; Grau, B.C.
\newblock {Adversarial Attacks as Near-Zero Eigenvalues in the Empirical Kernel
  of Neural Networks.} 
 2025.
 \newblock In Proceedings of the NeurIPS 2024 Workshop on Mathematics of Modern Machine Learning (M3L), Vancouver, BC, Canada, 14 December 2024.

\bibitem[Lundberg and Lee(2017)]{Lundberg2017unified}
Lundberg, S.; Lee, S.I.
\newblock A Unified Approach to Interpreting Model Predictions. \emph{arXiv} \textbf{2017}, arXiv:1705.07874.



\bibitem[Ge et~al.(2024)Ge, Patino, Todisco, and
  Evans]{ge2024explainingdeeplearningmodels}
Ge, W.; Patino, J.; Todisco, M.; Evans, N.
\newblock Explaining deep learning models for spoofing and deepfake detection
  with SHapley Additive exPlanations. \emph{arXiv} \textbf{2024}, arXiv:2110.03309.
  
  

\bibitem[Yu et~al.(2024)Yu, Chen, Leng, Chen, and Yi]{YU2024103720}
Yu, N.; Chen, L.; Leng, T.; Chen, Z.; Yi, X.
\newblock An explainable deepfake of speech detection method with spectrograms
  and waveforms.
\newblock {\em J. Inf. Secur. Appl.} {\bf 2024},
  {\em 81},~103720.
\newblock {\url{https://doi.org/10.1016/j.jisa.2024.103720}}.

\bibitem[Li et~al.(2025)Li, Ahmadiadli, and
  Zhang]{li2025surveyspeechdeepfakedetection}
Li, M.; Ahmadiadli, Y.; Zhang, X.P.
\newblock A Survey on Speech Deepfake Detection. \emph{arXiv} \textbf{2025}, arXiv:2404.13914.

\bibitem[Pomponi et~al.(2021)Pomponi, Scardapane, and
  Uncini]{Pomponi2021ProbabilisticConfidenceMultiExit}
Pomponi, J.; Scardapane, S.; Uncini, A.
\newblock A Probabilistic Re-Interpretation of Confidence Scores in Multi-Exit
  Models.
\newblock {\em Entropy} {\bf 2021}, {\em 24},~1.
\newblock {\url{https://doi.org/10.3390/e24010001}}.

\bibitem[Heidemann et~al.(2021)Heidemann, Schwaiger, and
  Roscher]{heidemann2021measuring}
Heidemann, L.; Schwaiger, A.; Roscher, K.
\newblock Measuring Ensemble Diversity and Its Effects on Model Robustness.
\newblock In Proceedings of the 1st International Workshop
  on Artificial Intelligence Safety (SafeAI 2021) Co-Located with AAAI 2021, CEUR-WS, {Virtually, 8 February} 2021; Volume 2916, pp. 65--73.

\bibitem[Lundberg et~al.(2019)Lundberg, Erion, Chen, DeGrave, Prutkin, Nair,
  Katz, Himmelfarb, Bansal, and Lee]{lundberg2019explainableaitreeslocal}
Lundberg, S.M.; Erion, G.; Chen, H.; DeGrave, A.; Prutkin, J.M.; Nair, B.;
  Katz, R.; Himmelfarb, J.; Bansal, N.; Lee, S.I.
\newblock Explainable AI for Trees: From Local Explanations to Global
  Understanding. \emph{arXiv} \textbf{2019}, arXiv:1905.04610.
  
  

\bibitem[Hajjouz and Avksentieva(2025)]{Hajjouz2025enhancing}
Hajjouz, A.; Avksentieva, E.
\newblock Enhancing and extending CatBoost for accurate detection and
  classification of DoS and DDoS attack subtypes in network traffic.
\newblock {\em Sci. Tech. J. Inf. Technol. Mech. Opt.} {\bf 2025}, {\em 25},~114--127.
\newblock {\url{https://doi.org/10.17586/2226-1494-2025-25-1-114-127}}.

\bibitem[Shen et~al.(2018)Shen, Pang, Weiss, Schuster, Jaitly, Yang, Chen,
  Zhang, Wang, Skerrv-Ryan, Saurous, Agiomvrgiannakis, and Wu]{8461368}
Shen, J.; Pang, R.; Weiss, R.J.; Schuster, M.; Jaitly, N.; Yang, Z.; Chen, Z.;
  Zhang, Y.; Wang, Y.; Skerrv-Ryan, R.;  et~al.
\newblock Natural TTS Synthesis by Conditioning Wavenet on MEL Spectrogram
  Predictions.
\newblock In Proceedings of the 2018 IEEE International Conference on
  Acoustics, Speech and Signal Processing (ICASSP), {Calgary, AB, Canada, 15--20 April} 2018; pp. 4779--4783.
\newblock {\url{https://doi.org/10.1109/ICASSP.2018.8461368}}.

\bibitem[Siuzdak(2024)]{Siuzdak2024Vocos}
Siuzdak, H.
\newblock Vocos: Closing the Gap Between Time-Domain and Fourier-Based Neural
  Vocoders for High-Quality Audio Synthesis.
\newblock In Proceedings of the International Conference on
  Learning Representations (ICLR), {Vienna, Austria, 7--11 May} {2024}
.


\bibitem[Pons et~al.(2021)Pons, Pascual, Cengarle, and Serrà]{9414913}
Pons, J.; Pascual, S.; Cengarle, G.; Serrà, J.
\newblock Upsampling Artifacts in Neural Audio Synthesis.
\newblock In Proceedings of the ICASSP 2021---2021 IEEE International
  Conference on Acoustics, Speech and Signal {Processing}
, {Toronto, ON, Canada, 6--11 June} 2021; \mbox{pp. 3005--3009}.
\newblock {\url{https://doi.org/10.1109/ICASSP39728.2021.9414913}}.

\bibitem[Tak et~al.(2021)Tak, weon Jung, Patino, Kamble, Todisco, and
  Evans]{tak21_asvspoof}
Tak, H.; weon Jung, J.; Patino, J.; Kamble, M.; Todisco, M.; Evans, N.
\newblock End-to-end spectro-temporal graph attention networks for speaker
  verification anti-spoofing and speech deepfake detection.
\newblock In Proceedings of the 2021 Edition of the Automatic Speaker
  Verification and Spoofing Countermeasures Challenge, {Online, 16 September} 2021; pp. 1--8.
\newblock {\url{https://doi.org/10.21437/ASVSPOOF.2021-1}}.

\bibitem[Zhang et~al.(2024)Zhang, Long, Cai, Yu, Shi, and
  Tan]{Zhang2024multislice}
Zhang, Q.; Long, Y.; Cai, H.; Yu, S.; Shi, Y.; Tan, X.
\newblock A multi-slice attention fusion and multi-view personalized fusion
  lightweight network for Alzheimer’s disease diagnosis.
\newblock {\em BMC Med. Imaging} {\bf 2024}, {\em 24}, 258.
\newblock {\url{https://doi.org/10.1186/s12880-024-01429-8}}.

\bibitem[Ben-Artzy and Schwartz(2024)]{benartzy2024attend}
Ben-Artzy, A.; Schwartz, R.
\newblock Attend First, Consolidate Later: On the Importance of Attention in
  Different LLM Layers. \emph{arXiv} \textbf{2024}, arXiv:2409.03621.
  
  

\bibitem[Tak et~al.(2021)Tak, Patino, Todisco, Nautsch, Evans, and
  Larcher]{9414234}
Tak, H.; Patino, J.; Todisco, M.; Nautsch, A.; Evans, N.; Larcher, A.
\newblock End-to-End anti-spoofing with RawNet2.
\newblock In Proceedings of the ICASSP 2021---2021 IEEE International
  Conference on Acoustics, Speech and Signal {Processing}, {Toronto, ON, Canada, 6--11 June} 2021; pp.
  6369--6373.
\newblock {\url{https://doi.org/10.1109/ICASSP39728.2021.9414234}}.

\bibitem[Rosello and Evans(2023)]{rosello2023conformer}
Rosello, V.; Evans, N.
\newblock A Conformer-Based Classifier for Variable-Length Utterance Processing
  in Anti-Spoofing.
\newblock In Proceedings of the INTERSPEECH, {Dublin, Ireland, 20--24 August} 2023; pp. 3632--3636.
\newblock {\url{https://doi.org/10.21437/Interspeech.2023-1627}}.

\bibitem[Wang et~al.(2024)Wang, Delgado, Tak, weon Jung, jin Shim, Todisco,
  Kukanov, Liu, Sahidullah, Kinnunen, Evans, Lee, and
  Yamagishi]{wang24_asvspoof}
{Wang, X.; Delgado, H.; Tak, H.; Jung, J.-W.; Shim, H.-J.; Todisco, M.;}
  Kukanov, I.; Liu, X.; Sahidullah, M.; Kinnunen, T.H.;  et~al.
\newblock {ASVspoof 5}: Crowdsourced speech data, deepfakes, and adversarial
  attacks at scale.
\newblock In Proceedings of the Automatic Speaker Verification Spoofing
  Countermeasures Workshop (ASVspoof 2024), {Kos, Greece, 31 August} 2024; pp. 1--8.
\newblock {\url{https://doi.org/10.21437/ASVspoof.2024-1}}.

\bibitem[D'Alterio et~al.(2025)D'Alterio, Neghina, Bestagini, and
  Tubaro]{dalterio2025attention}
D'Alterio, G.; Neghina, M.; Bestagini, P.; Tubaro, S.
\newblock Attention-based Mixture of Experts for Robust Speech Deepfake
  Detection.
\newblock In Proceedings of the IEEE International Workshop on
  Information Forensics and Security (WIFS), {Perth, WA, Australia, 1--4 December} {2025}
.


\bibitem[Tran et~al.(2025)Tran, Amsaleg, and Ducq]{tran2025multilevel}
Tran, H.M.; Amsaleg, L.; Ducq, E.
\newblock Multi-level SSL Feature Gating for Audio Deepfake Detection.
\newblock In \emph{Proceedings of the ACM International Conference on
  Multimedia Retrieval (ICMR), {Chicago, IL, USA, 30 June--3 July 2025}}; ACM:  {New York, NY, USA,} 
  2025.

\end{thebibliography}

\PublishersNote{}
\end{adjustwidth}
\end{document}